\documentclass{ifacconf}
\usepackage{float}
\usepackage{amsmath}
\usepackage{amssymb,bm}
\usepackage{comment}
\usepackage{amsfonts}
\usepackage{color}
\usepackage{lipsum}
\usepackage{romannum}
\usepackage{algorithm,algcompatible}
\usepackage{subcaption}
\usepackage{algpseudocode}
\usepackage{tikz}
\usepackage[pdftex]{pict2e}
\usepackage{graphicx}      
\usepackage{natbib}        
\usepackage{caption}        

\begin{document}
\begin{frontmatter}
\title{A Truthful Mechanism Design for Distributed Optimisation Algorithms in Networks with Self-interested Agents}


\author[First]{Tianyi Zhong}  
\author[Second]{David Angeli$^{*,}$} 

\address[First]{Control and Power Group, Electrical and Electronic Engineering Department, Imperial College, London, United Kingdom (e-mail: tianyi.zhong20@ imperial.ac.uk, d.angeli@imperial.ac.uk).}
\address[Second]{Dip. di Ingegneria dell’Informazione, University of Florence, Italy (e-mail: david.angeli@unifi.it)}

\begin{abstract}    
Enhancing resilience in multi-agent systems in the face of selfish agents is an important problem that requires further characterisation. This work develops a truthful mechanism that avoids self-interested and strategic agents maliciously manipulating the algorithm. We prove theoretically that the proposed mechanism incentivises self-interested agents to participate and follow the provided algorithm faithfully. Additionally, the mechanism is compatible with any distributed optimisation algorithm that can calculate at least one subgradient at a given point. 
Finally, we present an illustrative example that shows the effectiveness of the mechanism.
\end{abstract}


\end{frontmatter}

\section{Introduction}
Distributed optimisation has emerged as a critical framework in addressing complex, large-scale problems across various domains, including network design, machine learning, energy systems, and economics. In a distributed setup, agents need to make local decisions and negotiate with each other until an agreement is reached. The distributed nature of these algorithms offers several advantages, such as scalability, robustness, and privacy. These algorithms efficiently converge to a solution of the global problem if agents are trustworthy and fully cooperative. 
 However, it also introduces new challenges, particularly when agents are misbehaving in the algorithm. \cite{sundaram2018distributed} demonstrate that even a single malicious agent can drive consensus-based distributed optimisation algorithms to converge to any arbitrary value.

To mitigate the impact imposed by malicious agents, resilient distributed algorithms have been developed. Algorithms proposed by \cite{su2020byzantine}, \cite{zhang2023accelerated} focus on embedding a filtering process to achieve resilience. Our previous work \citep{angeli2023causal} involves agents checking and filtering their neighbours' gradient data to limit the effect of malicious agents to the system and recover convergence of the algorithm. \cite{dayi2024fast} propose to utilise inter-agent trust through a learning protocol. As discussed by \cite{sundaram2018distributed} in Theorem 4.4, a common challenge for these methods is that agents can simply alter their local cost functions and pretend to be regular agents in the algorithm. In general, these manipulations can hardly be detected. Hence, we are interested in designing an incentive mechanism to encourage agents not to adopt such a strategic manipulation.
 
There is a long tradition of using mechanism design techniques that focus on designing interaction and allocation rules to recover a socially (globally) optimal solution in the presence of strategic agents. The well-known mechanism that encourages agents to faithfully participate in the game is the Vickrey-Clarke-Groves (VCG) mechanism. It is implemented in the maximisation of social welfare for a smart grid by \cite{samadi2012advanced}. \cite{fu2012stochastic} combine VCG mechanism and conjectural pricing to regulate the virtualised wireless resources. The VCG mechanism involves monetary transfers to ensure truthfulness, however, the budget balance cannot be guaranteed. Nash mechanisms, on the other hand, preserve individual rationality and budget balance but come at the cost of losing dominant strategy incentives. Nash mechanisms are widely proposed for specific allocation problems. \cite{johari2009efficiency} present a limited communication VCG-like Nash mechanism for flow control problems, where the revealed utility functions are differentiable for every parameter. The differentiability assumption is removed by \cite{jain2010efficient}, in which a VCG-style Nash mechanism is designed for network resource allocation problems. A Nash mechanism is proposed by \cite{6125995} for a power allocation and spectrum sharing problem, however, the authors do not present an algorithm for the computation of the desired Nash equilibria (NE). These mechanisms are fragmented and case-by-case approaches. A general mechanism, achieving full Nash implementation if the agents' valuation functions are their private information, is proposed by \cite{sinha2014general} through analysing the dual optimisation problem. However, this work does not provide an algorithm that guarantees convergence to NE as well.

All the mentioned works focus on one-shot mechanisms and are not suitable in cases where actions unfold over time and agents have the opportunity to adjust their strategies based on new information. Recent works focus more on designing dynamic mechanisms. The dynamic VCG mechanism designed by \cite{bergemann2010dynamic} achieves ex-post incentive compatibility by assuming agent dynamics are coupled only through the action of the principal. \cite{garcia2015efficient} propose a dynamic truthful mechanism where the payment rule is designed under the assumption that agents’ valuations are independent. Inspired by this work, \cite{zhang2019efficient} explore the dynamic mechanism that is suitable to deal with asymmetric constraint information under the assumption that the influence of agent actions can be monitored. Using a surrogate optimisation approach, \cite{farhadi2018surrogate} propose a mechanism and introduce a tatonnement process for agents to converge to NE.

These works are developed with a focus on guaranteeing the desired properties of the mechanism and providing algorithms that converge to NE or GNE. Different from this focus, the motivation of this work is enhancing resilience of the distributed algorithm. In this work, we ask how to utilise mechanism design approaches to prevent strategic agents from altering their cost function and pretending to be regular agents in the algorithm. More related work considers the truthful implementation of distributed algorithms. 
\cite{parkes2004distributed} explore the distributed implementation of the VCG mechanism by proposing principles to guide the distribution of computation. This agenda has been advanced by \cite{petcu2006mdpop} which integrates mechanism design methods with one distributed optimisation algorithm and enables the algorithm to scale well in large-scale networks by removing the requirement to compute \(N+1\) (N denotes the number of agents) parallel optimisation problems. \cite{tanaka2015faithful} propose a general indirect mechanism design framework based on the VCG mechanism for faithful distributed algorithm implementations but lose the dominant strategy incentives.
 Considering that each agent has individual objective functions, \cite{zheng2017bridge} propose a gradient-based incentive mechanism within the framework of an alternating direction method of multipliers (ADMM) to impel truthful information reporting. 

These works model the agent as strategic and friendly, they neglect agents' strategic manipulation of other agents' payments (which only influences other agents' payoffs). This work models the agent as strategic and malicious but conservative and considers how they would act to benefit themselves and harm others at the same time.
We combine fully distributed optimisation algorithms and truthful mechanism design in the network with strategic agents. 
The only centralised control is to select outcomes and calculate payments. Our main contributions are as follows:
\begin{enumerate}
    \item We study a strongly convex social optimisation problem with the presence of strategic agents. Each agent is allowed to deviate from the proposed optimisation algorithm so as to decrease his own cost. The deviation is made by using an evaluation function that is not necessarily the same as his true cost function in the algorithm.
    \item We propose a mechanism that achieves asymptotic efficiency and asymptotic incentive compatibility, while also incentivising agents to participate under time-invariant sequence-independent or time-invariant sequence-dependent manipulation strategies (which will be formally defined later).
    \item The proposed mechanism is distributed in the sense that it is compatible with any gradient-based distributed optimisation algorithm. A central unit representing the authority is needed to select outcomes and calculate payments and is not involved in the distributed optimisation process. Payment calculation involves filtering gradient information and the filter acts offline, thus not affecting the overall convergence speed. 
\end{enumerate}
\begin{figure}[h]
\centering
\includegraphics[width=\columnwidth]{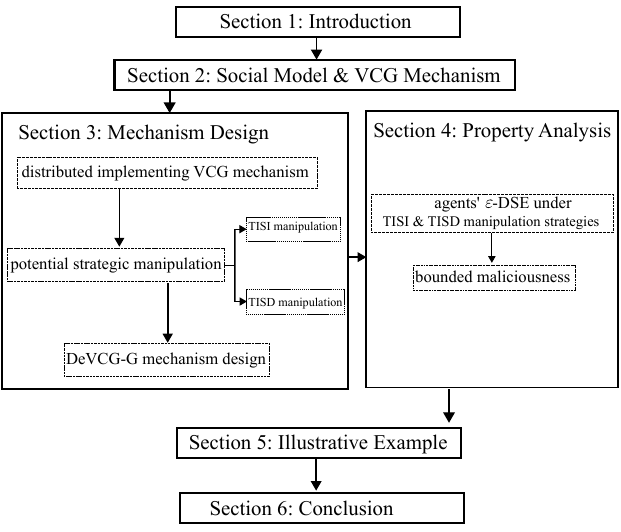}
\caption{\footnotesize Roadmap of the paper}
\label{figa}
\end{figure}
\section{Problem Formulation}
\subsection{Basic notation and notions}
We denote a tuple of functions \((f_1,f_2,\dots,f_N)\) as \((f_i,f_{-i})\) where \(f_{-i}=(f_1,\dots,f_{i-1},f_{i+1},\dots,f_N)\). The vertical concatenation of vectors \(x_i\in\mathbb{R}^n,i\in\mathcal{N}\) is denoted as \((x_i)_{i\in\mathcal{N}}\). Let \(x=(x_i)_{i\in\mathcal{N}}\in\mathbb{R}^{Nn}\), we use \([x]_i\in\mathbb{R}^n\) to represent the \(i\)-th portion \(x_i\) in the vector \(x\) and use \([x]_{-i}\) to represent \(((x_j)_{j\neq i})\). Specifically, in cases that \(x=(x_i)_{i\neq j}\in\mathbb{R}^{(N-1)n}\), \([x]_i\in\mathbb{R}^n\) refers to the portion contributed by agent \(i, i\neq j\). We use \(\partial v(x)\) to denote the subdifferential of \(v\) at a point x.
\subsection{Problem setup}
Consider a multi-agent system with a group of \(\mathcal{N} =\{1,2,\cdots,N\}\) agents. The communication between agents is modelled by a graph \(\mathcal{G}\). We are interested in the case where all agents are rational, selfish and possibly malicious.
\begin{align}
    &\min_x \sum_{i \in \mathcal{N}} f_i(x)\label{1}\\ 
   &{\rm s.t.}\quad x\in \mathcal{X} \notag
\end{align}
We make the following assumption on the cost functions and the constraint set.

\textbf{Assumption 1}: The cost functions \(f_i(x)\) are Lipschitz continuous for all \(i\in \mathcal{N}\) and are \(\mu\)-strongly convex such that the following holds for any \(x_1,x_2\in \mathcal{X}\):
\begin{align}
f_i(x_2)\geq f_i(x_1)+\nabla f_i(x_1)^T(x_2-x_1)+\frac{\mu}{2}\|x_2-x_1\|^2 
\end{align}
where \(\nabla f_i(x_1)\in\partial f_i(x_1)\) and the constraint set \(\mathcal{X}\subseteq  \mathbb{R}^n\) is assumed to be the interior of a convex and compact set.

Following the Lipschitz continuity of \(f_i\), we see the subgradients of each cost function \(f_i\) are bounded by the Lipschitz constant. There exists \(L\) such that for all \(i\in\mathcal{N}\):
\begin{align}
    \|\nabla f_i(x)\|\leq L, \quad  \nabla f_i(x)\in\partial f_i(x), x\in\mathcal{X}
\end{align}

Because there exists a point common to every set $\text{ri(dom} f_i),i\in\mathcal{N}$, where ri\((\cdot)\) denotes the relative interior of a set, let \(f(x)=\sum_{i\in\mathcal{N}}f_i(x)\), we have \(\nabla f(x) = \sum_{i\in\mathcal{N}}\nabla f_i(x),\nabla f_i(x)\in\partial f_i(x)\).

\textbf{Remark 1} Assumption 1 is a basic assumption for distributed convex optimisation algorithms. More assumptions on the cost function and constraints can be included when a specific optimisation algorithm is selected.

The strong convexity assumption ensures that social problem (\ref{1}) possess a unique minimiser.

\textbf{Assumption 2 (Selfish and Malicous Behaviour)}: We assume all agents are self-interested: they aim to maximise their utility without any regard for others. Meanwhile, we assume that agents are malicious: 
they tend to decrease other players' utility. 

Well-established distributed optimisation algorithms that solve problem (1) are two-fold: a state-based algorithmic model, and data transmission between agents. At each iteration, agents update their local states based on the algorithm dynamic and transmit data that depend on states.

We model the network algorithm dynamic as a mapping \(\sigma:\mathbb{N}\times\mathbb{R}^{Nn}\times\mathcal{V}^{N}\to\mathbb{R}^{Nn} \) whose inputs are the iteration step, the collection of appropriate agents' initialisation, and the collection of agents' cost functions while the output is the collection of agents' local estimation of the optimiser at a step \(k\):
\begin{align}
    x_{k} = \sigma(k,x_0,f)\label{4}
\end{align}
where \(x_k\in\mathbb{R}^{Nn}\) is the collection of all agent \(i\)'s estimation \(x^k_i\) and \(f\) denotes the collection of all local cost functions \(f_i,\forall i\in\mathcal{N}\). Also, \(\mathcal{V}\) is defined as:
\begin{align}
    \mathcal{V}=\{&v:\mathcal{X}\to \mathbb{R}:v \text{ satisfies Assumption 1}\}\cup \{\emptyset\}\label{5}
\end{align} 

Meanwhile, we model the data transmission process for agent \(i\in\mathcal{N}\) as a mapping \(s_i:\mathbb{R}^n\times\mathcal{V}\to\mathbb{R}^s\) whose inputs are the current step local state and his local cost function:
\begin{align}
    z^k_i = s_i(x^k_i,f_i)\label{6}
\end{align}
where \(z^k_i\in\mathbb{R}^s\) denotes the transmitted data from agent \(i\) at step \(k\).

In this work, we consider the situation where malicious agents can switch their local cost function to another function and pretend to be regular agents in the algorithm. We denote the switched cost function as an evaluation function. Note that this type of manipulation can hardly be detected and can drive the optimal result (outcomes) to an arbitrary point. Instead of arbitrarily driving the outcome, we assume that self-interested malicious agents have different preferences over possible outcomes and tend to their preference without any regard of others. 

Unlike malicious-resilient algorithms that focus on fault tolerance of processes (\ref{4}) and (\ref{6}), this paper considers problem (\ref{1}) as a game where agents can decide on using any evaluation function \(v_i\in\mathcal{V}\) to join the algorithm. Once the evaluation function \(v_i\) is chosen, the selected distributed algorithm is part of the game rule that participating agents must follow.

Problem (\ref{1}) is a one-shot game and evaluation functions that the designer would like every agent to adopt are their true costs \((f_i,f_{-i})\). However, this desire cannot be achieved as self-interested agents know that playing faithfully in the game will lead to an outcome that does not align with their interests.

\subsection{VCG mechanism}
One way to realign individual and social interests is to design incentive mechanisms. A mechanism \(\Gamma=(\mathcal{M},O(m),P(m))\) contains 1) a message space \(\mathcal{M}=\Pi_{i\in\mathcal{N}}\mathcal{M}_i\) including messages \(m=(m_i,m_{-i}), m_i\in\mathcal{M}_i\) that an agent \(i\in\mathcal{N}\) sends to the central authority; 2) an outcome function \(O:\mathcal{M}\to\mathbb{R}\) that determines the social outcome \(o\in\mathcal{X}\) based on the messages received from agents; and 3) Payment functions for each agent, \(P_i:\mathcal{M}\times\mathcal{X}\to\mathbb{R}\) which is used by the central authority to calculate each agent's payment after deciding the social outcome. Once a mechanism is in place, it induces a game \(G=(\mathcal{N},\mathcal{V}^{N},U)\) among strategic agents in set \(\mathcal{N}\), and \(U=(u_i,u_{-i})\) collects agents' payoff functions. 

We introduce the famous VCG mechanism. The message space \(\mathcal{M}_i\) for each agent \(i\in\mathcal{N}\) is represented by the set of all evaluation functions he could report:
\begin{align}
    \mathcal{M}_i=\mathcal{V}
\end{align}
The standard VCG scheme defines the social outcome \(o^*=O(v)\) and payment \(p_i=P_i(o^*,v), v=(v_i, v_{-i})\in\mathcal{M}\), for each agent \(i\in\mathcal{N}\) as:
\begin{align}
    o^*=O(v)&:={\rm arg}\min_x \sum_{i\in\mathcal{N}}v_i(x)\label{5}\\
    p_i=P_i(o^*, v)&:=\sum_{j\neq i}v_j(o^*)-\min_x \sum_{j\neq i}v_j(x)\label{9}
\end{align}
The VCG mechanism induces a one-shot game \(G=(\mathcal{N},\mathcal{V}^N, U)\). The agent’s payoff for reporting the evaluation function \(v_i\), receiving outcome \(o^*\) and payment \(p_i\) is \(u_i(v) =-f_i(O(v))-P_i(O(v),v) =-f_i(o^*)-p_i\). 

A dominant strategy for agent \(i\) is defined as a message \(v^*_i\in\mathcal{M}_i\), if for all possible messages of the other players \(v_{-i}\in\Pi_{j\neq i}\mathcal{M}_j\), the following holds:
\begin{align}
    u_i(v^*_i,v_{-i})\geq u_i(v_i,v_{-i}),\,\forall v_i\in\mathcal{M}_i
\end{align}
It is known that the dominant strategy of agent \(i\) in this game is \(v^*_i=f_i\), which results in the optimal social outcome \(o^*={\rm arg}\min_x \sum_{i\in\mathcal{N}}f_i(x)\).
\section{Truthful Mechanism Design}
\subsection{Distributed implementation of the VCG mechanism}
The VCG mechanism cannot be directly implemented in a distributed setup as agents are not willing to transmit their private cost function entirely to the central authority.

To restrict the message space to include only partial information about the evaluation function, the computation of (\ref{5}) and the second term in (\ref{9}) are replaced by \(N+1\) distributed optimisation processes. 
We denote the minimisation problem in (\ref{5}) as the social sequence and the minimisation problem in (\ref{9}) as sequence \(i\). The components of the game are:
\begin{enumerate}
    \item Agents: An agent can only communicate with his neighbours, and the inter-agent communication topology is described as \(\mathcal{G}\). 
    \item Strategies: A strategy profile of an agent \(i\) is \(s_i=(s_{i0},(s_{ij})_{j\neq i})\in S_i=S_{i0}\times\Pi_{j\neq i} S_{ij}\). The strategy for each sequence (\(s_{i0}\) for the social sequence and \(s_{ij}\) for sequence \(j\)) is a selection of the initial state and an evaluation function. The strategy space for social sequence is \({S}_{i0}\subseteq\mathbb{R}^{n}\times \mathcal{V}\) and \({S}_{ij}\subseteq\mathbb{R}^{n}\times \mathcal{V}\) for sequence \(j\neq i\).
    \item Evaluation functions: We use \(v_i\) to denote the evaluation function adopted by agent \(i\), based on which the actions and decisions can be calculated. Claiming no interest with \(v_i=\emptyset\) is a feasible choice, and the agent would agree on the determined social outcome.
    \item Decisions: We use \(x^{k_f}_i(v)\) and \(x^{k_f}_{ij}(v),j\neq i\) to denote the outcome decisions proposed by agent \(i\)  with the overall evaluation function profile \(v=(v_i,v_{-i})\). Specifically, the decisions of agent \(i\) are \(x^{k_f}_i(v)=[\sigma(k_f,x_0,v)]_i\) and \(x^{k_f}_{ij}(v)=[\sigma(k_f,[x_0]_{-j},v_{-j})]_i\), where \([x_0]_{-j}\) denotes the collection of local initial states without agent \(j\). Agent \(i\)'s decisions \(x^{k_f}_i,x^{k_f}_{ij}\) are subject to a commonly known constraint set \(\mathcal{X}\subseteq\mathbb{R}^n\). If an agent \(i\) decides not to join the induced game, he does not make any decisions \(x^{k_f}_i=x^{k_f}_{ij}=\emptyset,\forall j\neq i\).
    \item Actions: An action is the transmitted data in the optimisation process. An action of agent \(i\) at step \(k\) with evaluation function \(v_i\) in the social sequence is \(z^k_i\) (as defined in (\ref{6}) with \(f_i\) be replaced by \(v_i\)) and \(z^k_{ij}\) in sequence \(j\).
\end{enumerate}

\begin{figure}[h]
\centering
\includegraphics[width=\columnwidth]{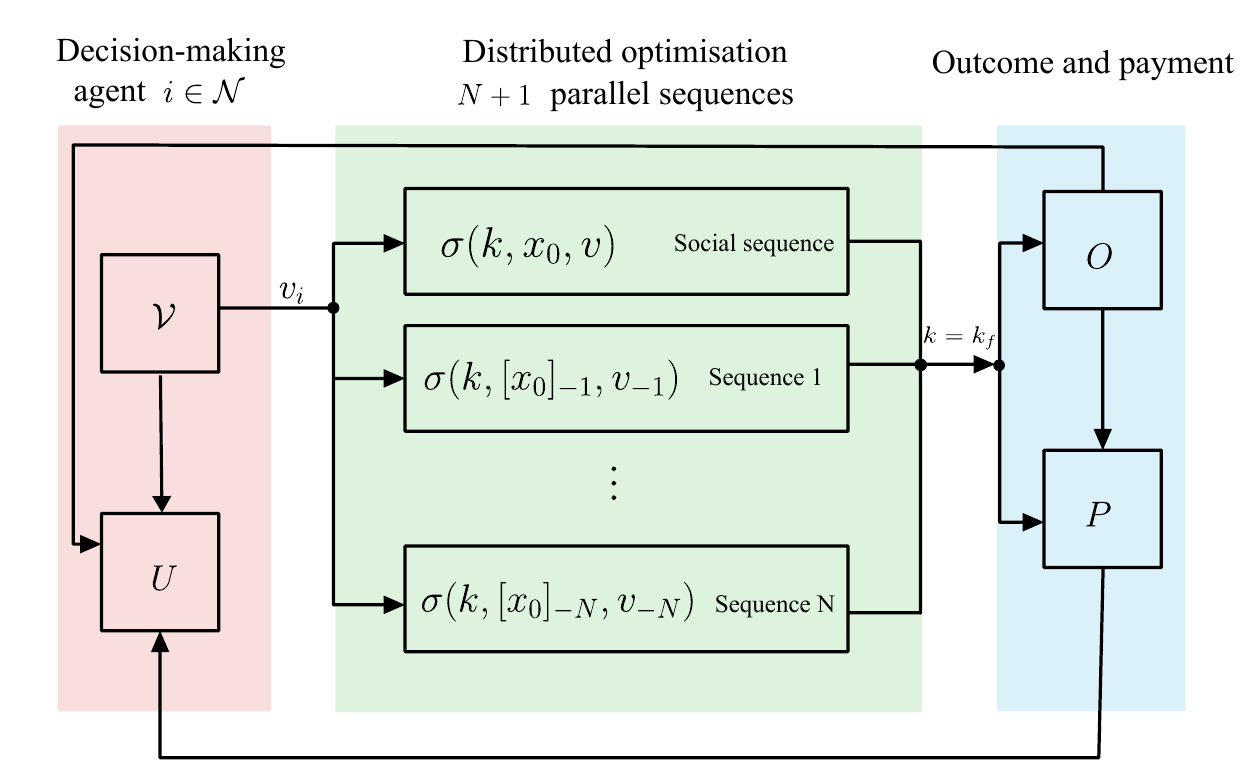}
\caption{ Schematic of the network with \(N\) agents joining the mechanism. The evaluation function space \(\mathcal{V}\) and agent's payoff function \(U\) (represented in pink boxes) define the game. The outputs of parallel algorithms \(\sigma\) define the message space. The message space, outcome mapping \(O\) and payment mapping \(P\) (represented in the blue box) define the mechanism.}
\label{fig0}
\end{figure}

Figure \ref{fig0} shows the structure of the network with \(N\) agents and a distributed mechanism. The detailed distributed VCG mechanism is shown in Mechanism 1. An agent \(i\) transmits messages to the central authority at two stages. For those \(i\in\mathcal{I}=\{1,\cdots,I\}\subseteq\mathcal{N}\) whose evaluation function is not empty \(v_i\neq \emptyset\): 1) when \(I+1\) distributed optimisations finish, the agent transmits the decision made in the social sequence and the decision made in sequences \(j\neq i\); and 2) after receiving the social outcome and sequences \(j\)s' outcomes from the central authority, the agent evaluate these outcomes according to their evaluation functions and transmits the results to the central authority.

The game induced by the distributed mechanism is denoted as \(G=(\mathcal{I},\mathcal{V}^I, k_f, U)\), where \(\mathcal{I}\) denotes the set of agents whose evaluation function is not empty, 
\(k_f\) is the maximum iteration step and \(U\) is the payoff mapping. 

Note that agents are also given the option of not participating in the game. In this case, agent \(i\)'s evaluation function is \(\emptyset\) and he does not have the ability to influence the social outcome and will not be charged an additional payment by the mechanism. The agent \(i\) agrees with the outcome determined by those who join the game. However, if all agents quit the game, the central authority will charge them a sufficiently large payment.

\textbf{DeVCG Game:} The DeVCG Mechanism induces
a DeVCG Game where each agent \(i\in\mathcal{N}\) selects an evaluation function \(v_i\in\mathcal{V}\) before joining the algorithm. In particular, quitting agent \(i\)'s strategies is \(v_{i}=\emptyset,i\in\mathcal{N}\setminus\mathcal{I}\). After \(k_f\) steps, each agent simultaneously submits decisions \(x^{k_f}_i(v)\) and \(x^{k_f}_{ij}(v), \forall j\neq i\) to the central authority:
\begin{subequations}
\label{x dynamic}
\begin{align}
    x^{k_f}_i(v)&=[\sigma(k_f,x_0,v)]_i\\
    x^{k_f}_{ij}(v)&=[\sigma(k_f,[x_0]_{-j},v_{-j})]_i 
\end{align}
\end{subequations}
We use the notation \(m_i(v,k_f)\) to represent agent \(i\)'s message when the evaluation profile is \(v\) and the maximum iteration step is \(k_f\). 
The outcomes \(o^*(m,k_f)\) and \(o_i(m_{-i},k_f)\) are selected by the central authority as the component-wise median of \(\{x^{k_f}_i\}_{i\in\mathcal{I}}\) and \(\{x^{k_f}_{ji}\}_{j\neq i}\), respectively.
\begin{subequations}
\label{o selection}
\begin{align}
       o^*(m,k_f)&=Median(x^{k_f}_i)_{i\in\mathcal{I}}\\
       o_i(m_{-i},{k_f})&=Median(x^{k_f}_{ji})_{j\neq i}
\end{align}
\end{subequations}
The decided outcomes are sent back to each agent. The notations \(o^*(m,k_f)\) and \(o_i(m_{-i},k_f)\) characterise the dependence of outcomes on the message and the maximum iteration step. 

Specifically, the message has the following form:
\begin{align}
    m_i(v,k_f)=(x^{k_f}_i,(x^{k_f}_{ij})_{j\neq i})\times(v_i(o^*), v_i(o_j)_{j\neq i})\label{12}
\end{align}
where \(o^*\) and \(o_j\) are used to simplify notations. Consider a decided evaluation profile \(v_{-i}\in\mathcal{V}^{N-1}\) for  other agents \(j\neq i\) and the total iteration step \(k_f\), the space of all possible messages from agent \(i\)'s perspective is:
\begin{align}
    \textbf{M}(v_{-i},k_f)&=\bigcup_{v_{i}\in\mathcal{V}}\Bigl\{\bigl(m_i(v_i,v_{-i},k_f),m_{j}(v_i,v_{-i},k_f)_{ j\neq i}\bigr) \notag\\
    &\text{ as in } (\ref{12}) :\,x^{k_f}_i,x^{k_f}_{ij},x^{k_f}_j,x^{k_f}_{jp}  \text{ are as in } (\ref{x dynamic})\notag\\
    & \text{ and }o^*,o_i,o_j \text{ are as in } (\ref{o selection}),\forall j,p\neq i\Bigr\}
    \label{message space}
\end{align}
The agent \(i\in\mathcal{I}\) decides the budget proposal as: 
\begin{align}
\tau_{i}(m,k_f) &= (v_i(o^*),v_i(o_j)_{j\neq i})\label{13}
\end{align} 
In particular, to clarify the notation as used in (\ref{11}), we denote \((\tau_{i}(m,k_f))_0 = v_i(o^*)\) and \((\tau_{i}(m,k_f))_j=v_i(o_j)\). Upon receiving the budget proposal, the central authority determines payment \(p_i(m,k_f)\) for each agent \(i\in\mathcal{I}\) and \(p_i=0\) for \(i\in\mathcal{N}\setminus\mathcal{I}\).

Strategic agents aim at selecting an evaluation function \(v_i\in\mathcal{V}\) to maximise their own payoff \(u_i(o^*,p_i)=-f_i(o^*)-p_i\) in this game.\hfill\(\square\)

\begin{algorithm}
\caption{Distributed VCG Mechanism (DeVCG)}
\begin{algorithmic}[1]
    \State \textbf{Initialisation}
    
     The central authority publishes the distributed negotiation algorithm (whose dynamic is (\ref{4})), the termination tolerance \(\varepsilon\) and the maximum iteration step \(k_f\). It is guaranteed by the central authority that the tolerance is satisfied at a step \(k\leq k_f\).
    \State \textbf{The message space \(\mathcal{M}=\Pi_i \mathcal{M}_i,\,\mathcal{M}_i\subset\mathcal{X}^I\times \mathbb{R}^I\)} 
    
     Each agent \(i\in\mathcal{N}\) submits a message \(m_i\in\mathcal{M}_i\) to the central authority:
     \begin{subequations}
     \label{10}\begin{align}
     {m}_i:=&{m}^1_i\times {m}^2_i\\
     {m}^1_i:=&\bigl\{\bigl(x^{k_f}_i, (x^{k_f}_{ij})_{j\neq i}\bigr)\bigr\}\\
     {m}^2_i:=&\{\tau_{i}(m^1,k_f)\}
     \end{align}
     \end{subequations}
     where \(x^{k_f}_i\) and \(x^{k_f}_{ij}\) are agent \(i\)'s decisions for the social outcome and sequence \(j\)'s outcome, respectively. The collection of local decisions are calculated using algorithm (\ref{4}), and \(\tau_{i}\) is agent \(i\)'s budget proposal which depends on \(m^1=(m^1_i,m^1_{-i})\).
    \State \textbf{Social \(\&\) sequences outcome \(o^*,o_i\in\mathbb{R}^n,i\in\mathcal{I}\)} 

   The social outcome \(o^*(m,k_f)\) is taken as the component-wise median of \(\{x^{k_f}_i\}_{i\in\mathcal{I}}\) and sequence \(i\)'s outcome \(o_{i}(m_{-i},k_f)\) is taken as the median of \(\{x^{k_f}_{ji}\}_{j\neq i}\):
   \begin{subequations}
   \begin{align}
       o^*(m,k_f)&=Median(x^{k_f}_i)_{i\in\mathcal{I}}\\
       o_i(m_{-i},{k_f})&=Median(x^{k_f}_{ji})_{j\neq i}
   \end{align}
   \end{subequations}

    \State \textbf{Payment \(p_i\in\mathbb{R},i\in\mathcal{N}\)} 

     The payment for each agent \(i\in\mathcal{I}\) is calculated as:
     \begin{align}
         p_i(m,k_f):=\sum_{j\neq i}(\tau_{j}(m,k_f))_0-\sum_{j\neq i}(\tau_{j}(m,k_f))_i\label{11}
     \end{align}
     where \((\tau_{j}(m,k_f))_0\) is the portion related to the social sequence and \((\tau_{j}(m,k_f))_i\) is the portion related to sequence \(i\).

     The payment for each agent \(i\in\mathcal{N}\setminus\mathcal{I}\) is \(p_i=0\).
     If all agents quit the game, the payment for each agent \(i\in\mathcal{N}\) is \(\bar{p}\), where \(\bar{p}<+\infty\) is positive and sufficiently large.
\end{algorithmic}
\end{algorithm}

\textbf{Remark 1.} It is important to note that the strategic choice available to each agent is their evaluation function, which indirectly influences the outcome of the mechanism. In contrast, the outcome itself is determined directly by the message profile received by the central authority. Even if an agent keeps their evaluation function fixed, their transmitted messages can still change in response to changes in other agents' evaluation functions due to the dynamics of the distributed algorithm. Based on this interdependence feature, we define the dominant strategy for each agent with respect to the evaluation function they choose, while the dominant strategy equilibrium of the game is defined in terms of the resulting message profile.

\textbf{Definition 1 (Dominant Strategy).} The dominant strategy for each agent \(i\in\mathcal{N}\) in the DeVCG game is defined as an evaluation function \(v^*_i\in\mathcal{V}\) such that for any \(v_{-i}\in\mathcal{V}^{N-1}\), the corresponding message profile \(\hat{m}^*=(\hat{m}^*_i,\hat{m}^*_{-i})\in\textbf{M}(v_{-i},k_f)\) fulfils the following inequality for all \(m=(m_i,m_{-i})\in\textbf{M}(v_{-i},k_f)\):
\begin{align}
    &u_i\left(o^*(\hat{m}^*,k_f),p_i(\hat{m}^*,k_f)\right)\geq u_i\left(o^*(m,k_f),p_i(m,k_f)\right) \label{dominant ineq}
\end{align}
where \(o^*(\hat{m}^*,k_f)=Median(x^{k_f}_i)_{i\in\mathcal{I}}\) and \(o^*(m,k_f)=Median(\hat{x}^{k_f}_{i})_{i\in\mathcal{I}}\). With an appropriate collection of initial states \(x_0\), the collection of \(x^{k_f}_i\) and \(\hat{x}^{k_f}_{i}\) are calculated as:
\begin{subequations}
\begin{align}
    x_{k_f}&=\sigma({k_f},x_0,v_i^*,v_{-i})\\    \hat{x}_{k_f}&=\sigma({k_f},x_0,v) 
\end{align}
\end{subequations}

We see agent \(i\)'s dominant strategy is his true cost function \(f_i\) and the corresponding messages in the DeVCG game with any \(v_{-i}\in\mathcal{V}^{N-1}\) is the following:
\begin{align}
m_i^*(f_i,v_{-i},+\infty)=(x^*,(x_{j}^*)_{j\neq i})\times(f_i(x^*),f_i(x^*_j)_{j\neq i})
\end{align}
where \(x^*=\text{arg}\min_{x\in\mathcal{X}}\{f_i(x)+\sum_{j\neq i}v_j(x)\}\) and \(x_j^*=\text{arg}\min_{x\in\mathcal{X}}\{f_i(x)+\sum_{p\in\mathcal{N}\setminus\{i,j\}}v_p(x)\}\).
Note that this dominant strategy can only be achieved when \(k_f\to+\infty\).

\textbf{Remark 2.} It is important to note that the comparison between \(\hat{m}^*=(\hat{m}^*_i,\hat{m}^*_{-i})\in\textbf{M}(v_{-i},k_f)\) and an arbitrary message profile \(m=(m_i,m_{-i})\in\textbf{M}(v_{-i},k_f)\) does not imply that the other agents have changed their evaluation functions. Rather, the difference arises solely from agent \(i\)'s change of evaluation function \(v_i\), and the associated deviation of messages. The evaluation functions of the other agents remain unchanged. Moreover, \(\hat{m}^*_{-i}\) should not be interpreted as the messages resulting from other agents playing their own dominant strategies; rather, they represent the messages that those agents would transmit in response to agent \(i\)'s dominant strategy, given their fixed evaluation functions.

Since the social outcome is determined by the result of a distributed algorithm, terminating the process after a finite number of steps will yield only an approximate solution and the condition \eqref{dominant ineq} can hardly be fulfilled in finite time. We employ the approximation notion that allows the dominant strategy inequality condition to be satisfied up to a small \(\varepsilon\), i.e., the agent is satisfied with the outcome at where his payoff is \(\varepsilon\) away from optimal.

\textbf{Definition 2 (\(\varepsilon\)-Dominant Strategy Equilibrium).}  An \(\varepsilon\)-dominant strategy equilibrium (\(\varepsilon\)-DSE) of the DeVCG game for any given \(\varepsilon\in(0,1)\) is defined as a message profile \(m^*=(m^*_i,m^*_{-i})\in\bigcap_{i\in\mathcal{N}}{\textbf{M}}(v^*_{-i},k_f)\) corresponding to an evaluation function profile \(v^*=(v^*_i,v^*_{-i})\in\mathcal{V}^N\) that collects all agents' dominant strategy. For all \(m=(m_i,m_{-i})\in{\bigcup_{i\in\mathcal{N}}\textbf{M}}(v^*_{-i},k_f)\) and all \(i\in\mathcal{N}\), there exists a finite step \(k\) and the following holds for any \(k_f\geq k\):
\begin{align}
    &u_i(o^*(m^*,{k_f}),p_i(m^*,{k_f}))+\varepsilon \label{ep dominant ineq}\\
    &\geq u_i(o^*(m,{k_f}),p_i(m,{k_f}))\notag
\end{align}
where \(o^*(m^*,k_f)=Median(x^{k_f}_i)_{i\in\mathcal{I}}\) and \(o^*(m,k_f)=Median(\hat{x}^{k_f}_{i})_{i\in\mathcal{I}}\). With an appropriate collection of initial states \(x_0\), the collection of \(x^{k_f}_i\) and \(\hat{x}^{k_f}_{i}\) are calculated as:
\begin{subequations}\label{23}
\begin{align}
    x_{k_f}&=\sigma({k_f},x_0,v^*)\\    
    \hat{x}_{k_f}&=\sigma({k_f},x_0,v_i,v_{-i}^*) 
\end{align}
\end{subequations}

\textbf{Remark 3.} To avoid ambiguity regarding condition \eqref{ep dominant ineq}, which is derived from \eqref{23}, we clarify that the term “dominant” here primarily refers to the fact that for any \(v_{-i}\in\mathcal{V}^{N-1}\), there exists a message profile \((\hat{m}^*_i,\hat{m}^*_{-i})\in \textbf{M}(v_{-i},k_f)\) induced by agent \(i\)'s dominant strategy \(v^*_i\) such that:
\begin{align}
    &u_i(o^*(\hat{m}^*,{k_f}),p_i(\hat{m}^*,{k_f}))+\varepsilon \label{24}\\
    &\geq u_i(o^*(m,{k_f}),p_i(m,{k_f}))\notag
\end{align}
holds for all \((m_i,m_{-i})\in \textbf{M}(v_{-i},k_f)\). Note, however, that the “equilibrium” referenced in this context corresponds to the message profile resulting when all agents play their dominant strategies. Therefore, the formulation in \eqref{23} should be distinguished from the conventional definition of a Nash equilibrium in game theory.

\textbf{Definition 3 (\(\varepsilon\)-Incentive Compatibility).} A distributed mechanism \(\Gamma_{k_f} = (\mathcal{M},O(m,{k_f}),P(m,{k_f}))\) 
is \(\varepsilon\)-incentive compatible (\(\varepsilon\)-IC) for some given \(\varepsilon\in(0,1)\) if an \(\varepsilon\)-DSE of its induced game is \(m^*=(m^*_i,m^*_{-i})\):
\begin{align}
m^*_i=(o^*,(o_j)_{j\neq i})\times(f_i(o^*),f_i(o_j)_{j\neq i})
\end{align}
where \(o^*=Median(x^{k_f}_i)\) and \(o_j=Median(x^{k_f}_{ij})\). With an appropriate collection of initial states \(x_0\), the collection of \(x^{k_f}_i\) and \(x^{k_f}_{ij}\) are calculated as:
\begin{subequations}
\begin{align}
    x_{k_f}&=\sigma({k_f},x_0,f_i,f_{-i})\qquad(\text{social problem})\\
    [x_{k_f}]_{-j}&=\sigma({k_f},[x_0]_{-j},f_{-j}) \,\,(\text{problem without }j)
\end{align}
\end{subequations}

\textbf{Definition 4 (Asymptotic Incentive Compatibility).} A sequence of distributed mechanism \(\{\Gamma_{k}\}\) is said to be asymptotic incentive compatible if for every \({k}\) sufficiently large, \(\Gamma_{k}\) is \(\varepsilon_k\)-IC and \(\lim_{k\to+\infty}\varepsilon_k=0\).

\textbf{Definition 5 (Asymptotic Efficient).} A sequence of distributed mechanism \(\{\Gamma_{k}\} \) is asymptotic efficient if for every \(\Gamma_k\) with \(k\) sufficiently large, there exist \(\varepsilon_k\in(0,1)\) and an \(\varepsilon_k\)-DSE of its induced game \(m^*=(m^*_i,m^*_{-i})\) 
fulfills:
\begin{align}
    &\sum_{i\in\mathcal{N}}f_i(o^*(m^*,k))\leq \sum_{i\in\mathcal{N}}f_i(x)+\varepsilon_k,\forall x\in\mathcal{X}
\end{align}
and \(\lim_{k\to+\infty}\varepsilon_k=0.\)

Directly from Definitions 2 -- 5, we obtain the following theorem.

\textbf{Theorem 1.} Let Assumptions 1 and 2 hold. Given any tolerance \(\varepsilon\in(0,1)\), there exists \(k\) such that DeVCG mechanism \(\Gamma_{k}\) is \(\varepsilon\)-IC and the sequence \(\{\Gamma_{k_f}\}_{k_f\geq k}\) is asymptotic efficient.
 
By distributing the VCG mechanism, agents begin interacting with one another, enabling them to adopt a sequence of manipulation strategies instead of committing to a single evaluation function prior to the mechanism's initialisation. As a result, the strategy space would be larger than the one in the DeVCG game.
Consequently, the desirable properties of the DeVCG mechanism may no longer hold. This motivates us to explore the potential manipulation strategies agents might adopt and the corresponding effects that could arise.

\subsection{Malicious and Strategic manipulation}
Two types of manipulation strategies can influence the optimisation process: time-invariant (TI) manipulations and time-varying (TV) manipulations. Given that there are \(I+1\) distributed optimisation problems, each manipulation type can be further categorised into two subtypes based on agents' interaction with the mechanism: sequence-independent (SI) manipulation strategies and sequence-dependent (SD) manipulation strategies. In this paper, we focus on the time-invariant manipulation strategies.

\textbf{Definition 6 (TISI Manipulation Strategy).} We say that an agent \(i\in\mathcal{N}\) plays according to a time-invariant sequence-independent manipulation strategy if he selects one evaluation function \(v_i\) from \(\mathcal{V}\) in all sequences prior to the initialisation of the mechanism. In this case, we identify the evaluation function space for each agent as \(\mathcal{V}_i=\mathcal{V}\).

\textbf{Definition 7 (TISD Manipulation Strategy).} We say that an agent \(i\in\mathcal{N}\) plays according to a time-invariant sequence-dependent manipulation strategy if he selects different evaluation functions in different sequences from \(\mathcal{V}\). Specifically, agent \(i\) selects \(v_{i}\in\mathcal{V}\) in the social sequence and selects \(v_{ij}\) from \(\mathcal{V}\) in sequence \(j\), \(\forall j\neq i\). This selection is done prior to the initialisation of the mechanism. The evaluation function space for each agent is \(\mathcal{V}_i=\mathcal{V}^{I}\).

It is important to understand why agents have an incentive to adopt such manipulation strategies. In games induced by VCG-based mechanisms, an individual agent \(i\)'s payoff function depends on other agents' strategies in sequence \(i\). A malicious and strategic agent \(j\neq i\) could choose sequence-dependant manipulation strategies to impose a large negative value in sequence \(i\) to significantly reduce agent \(i\)'s payoff, thereby discouraging other agents from participating in the game. As a result, the social outcome would be determined solely by agent \(j\).

\textbf{Lemma 1.} Given a tolerance \(\varepsilon\in(0,1)\) that is satisfied by iterating \(k_f\) steps in the distributed algorithm, the pure strategy Nash equilibria of the DeVCG game with agents using TISD manipulation strategies are \(m^*(\bm{v}_i,\emptyset,k_f), i\in\mathcal{N}\), where \(\bm{v}_i=(f_i,(v_{ij})_{j\neq i}),v_{ij}\in\mathcal{V}\), \(v_{ij}(x)\) is negative and sufficiently large for all \(x\in\mathcal{X}\).

\textbf{Proof.} Let all agents \(i\in\mathcal{N}\) be strategic and malicious in a way that \(v_{ij}(x)<-\bar{p}/(N-1), \forall x\in\mathcal{X}\) (\(\bar{p}\) is the same as in Mechanism 1) holds for all sequences \(j\neq i\). Let \(o^*\) be the social outcome and \(o_{i}\) be sequence \(i\)'s outcome. The payoff of individual agent \(i\in\mathcal{N}\) is:
\begin{align}
u_i=
\begin{cases}
    &-f_i(o^*),\,\,\text{only one agent join the game }\\
 & -\bar{p},\qquad \text{ all agents quit the game}\\
    &-f_i(o^*)-\sum_{j\neq i}v_{j}(o^*)+\sum_{j\neq i} v_{ji}(o_{i}), {\rm otherwise}
\end{cases}\notag
\end{align}
As \(\bar{p}\) is sufficiently large, we see the set of pure Nash equilibria entails a single agent \(i\in\mathcal{N}\) joining the DeVCG game and his evaluation function profile is \(\bm{v}_i=(f_i,(v_{ij})_{j\neq i})\), where \(v_{ij}(x)\) is negative and sufficiently large for all \(x\in\mathcal{X}\).\hfill\(\square\)

Lemma 1 shows that the DeVCG mechanism is inefficient under TISD manipulation strategies.

\subsection{DeVCG-G mechanism and its induced game}
To limit the maliciousness of agents, the mechanism needs to establish a direct linkage between agents' behaviour in all sequences and their own payoff. This is done by checking the consistency of agents' actions across all sequences. As the mechanism targets gradient-based distributed algorithms, the central authority checks the gradient-state consistency at each step along all sequences.

\textbf{Definition 8 (Gradient-state Consistency).} Given a gradient-based distributed algorithm and agent \(i\)'s corresponding gradient-state profile \((g^k_i,x^k_i)\) and \((g^k_{ij},x^k_{ij}),\forall j\neq i\) at each step in all sequences, we say they are consistent if there exists an evaluation function \(v_i\in\mathcal{V}\) such that for all \(i,j\in\mathcal{I}\) and all \(k\geq 0\):
\begin{subequations}
\begin{align}
    g^k_i&\in\partial v_i(x^k_i)\\
    g^k_{ij}&\in\partial v_i(x^k_{ij})
\end{align}
\end{subequations}

We introduce the proposed DeVCG-G mechanism that combines the DeVCG mechanism with a resilient gradient filter. This filter, proposed by \cite{angeli2023causal}, is designed to verify whether a sequence of gradient and state data corresponds to an existing but unknown convex function. If the data violates the convexity assumption, the filter takes appropriate actions to minimally adjust the received gradient data, thereby recovering convexity. This filter is used to check if the evaluation functions of an agent in all sequences and all steps correspond to the same convex function.

We employ \(I\) data filtering nodes within the central unit, denoted as \(\mathcal{F}_i,i\in\mathcal{I}\). Each agent \(i\in\mathcal{I}\) sends the messages \(m^1_i(k)\) and \(m^2_i(k)\) (to be introduced in the DeVCG-G mechanism and game) to \(\mathcal{F}_i\). The central authority randomly decides a step \(k_s<k_f\) at which \(\mathcal{F}_i\) starts filtering \(\{m^2_i(k)\}_{k_s\leq k\leq k_f}\) and uses the filtered information to calculate agent \(i\)'s payment. However, the filtered information will be kept private within the central authority. Adopting \(k_s> 0\) is to reduce the computation and memory load, but ideally, the filter would be activated from step 0.

The central authority first casually combines the received gradient  \(\{g^t_i,(g^t_{ij})_{j\neq i}\}\) and state \(\{x^t_i,(x^t_{ij})_{j\neq i}\}, t\leq k\) to obtain \(\{\xi^t_i\}_{0\leq t\leq I(k+1)-1}\) and \(\{\eta^t_i\}_{0\leq t\leq I(k+1)-1}\). Specifically, for any \(t\in[0,k]\), \(\xi^{It}_i=g^t_i\), \(\xi^{It+j}_i = g^t_{ij}, 1\leq j\leq i-1\) and \(\xi^{It+j}_i=g^t_{ij+1},i\leq j\leq I-1\). Then, \(\mathcal{F}_i,\forall i\in\mathcal{I}\) filters the gradient data at \(t\in[Ik,I(k+1)-1]\):
\begin{align}
    &\Tilde{\xi}_i^{t}={\rm arg}\min_{\xi\in \mathbb{R}^n}\|\xi-\xi_i^{t}\|^2 \label{17c}\\ 
    &{\rm s.t.}\,(\xi^T(\eta_i^{\tau}-\eta_i^{t})+F_i^{\tau,m}(t-1)+(\Tilde{\xi}_i^{m})^T(\eta_i^{t}-\eta_i^{m})\leq 0 \notag\\
    &\qquad\forall \tau,m \in \{1,2,\dots,t-1\} \notag
\end{align}
where \(F_i^{1,1}(0):=0,\forall i \in\mathcal{I}\):
\begin{align}
    &F_i^{\tau,t}(t)=\max_{m\in\{1,\dots,t-1\}}F_i^{\tau,m}(t-1)+(\Tilde{g}_i^m)^T(\eta_i^{t}-\eta_i^m)\notag\\
    &F_i^{t,m}(t)=\max_{\tau\in\{1,\dots,t-1\}}(\Tilde{g}_i^{t})^T(\eta_i^\tau-\eta_i^{t})+F_i^{\tau,m}(t-1)\notag\\
    &F_i^{\tau,m}(t)=\max\{F_i^{\tau,m}(t-1),F_i^{\tau,t}(t)+F_i^{t,m}(t)\}\notag
\end{align}
 \begin{figure}
 \centering
\centerline{\includegraphics[width=1\columnwidth]{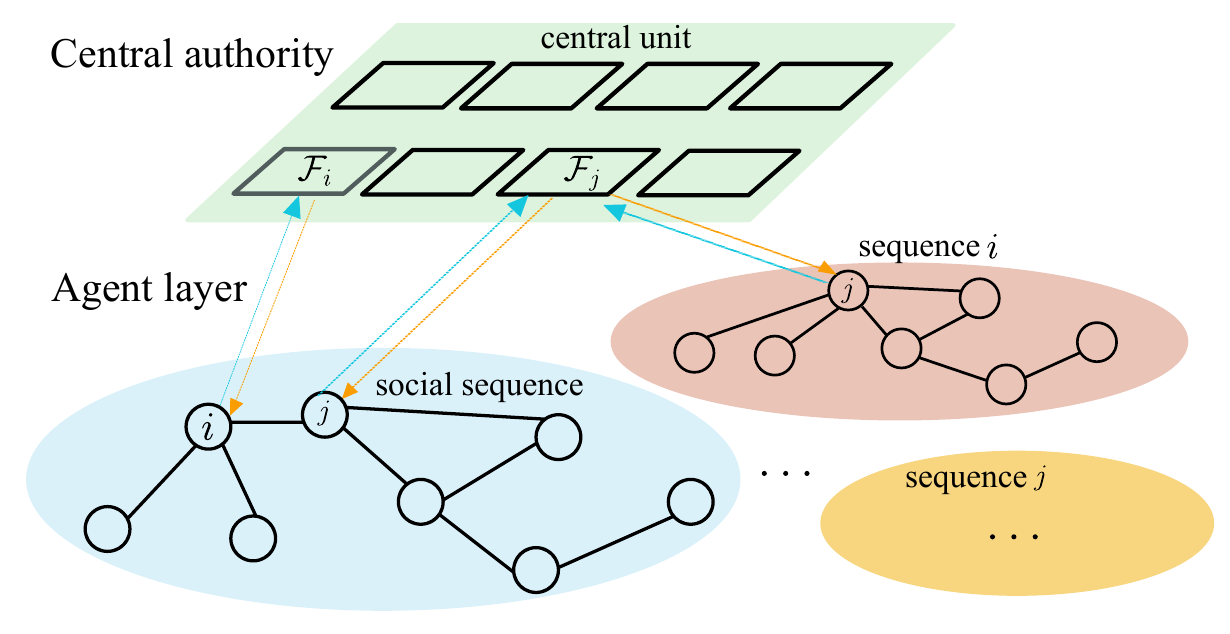}}
\caption{The upper layer presented in green is the central unit that contains \(I\) filtering nodes \(\mathcal{F}_i\). The lower networks are inter-agent networks in \(I+1\) different sequences. Agents transmit algorithmic variables to distributively find the solution of the optimisation problems.}
\label{fig3}
\end{figure}

\begin{figure}
 \centering
\centerline{\includegraphics[width=0.9
\columnwidth]{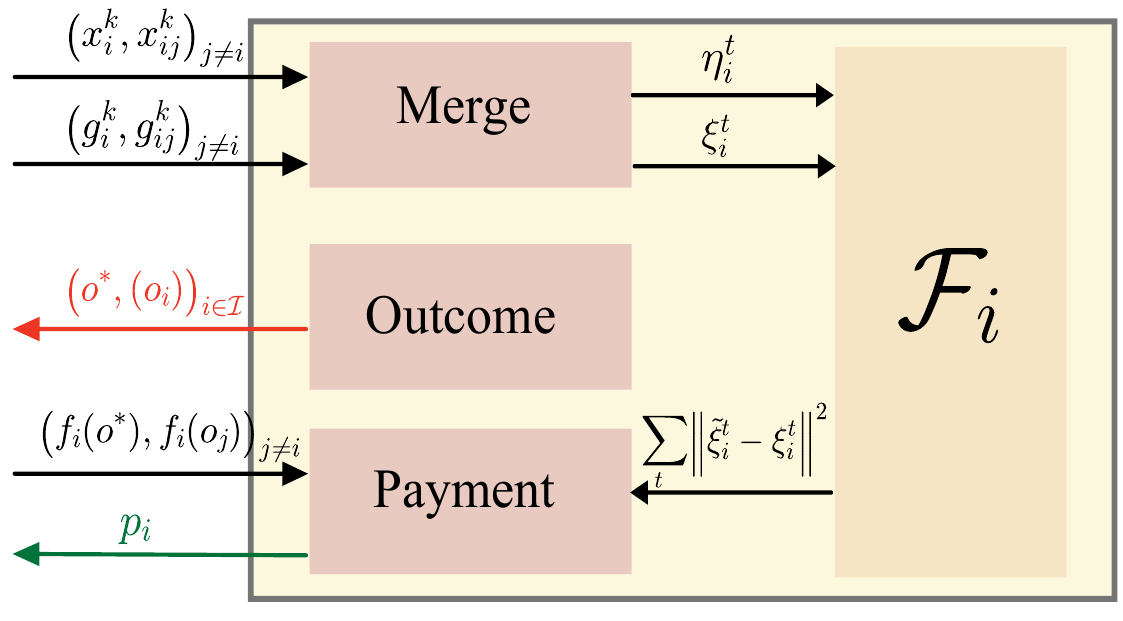}}
\caption{Central unit structure}
\label{fig3-1}
\end{figure}

\begin{figure}
 \centering
\centerline{\includegraphics[width=1\columnwidth]{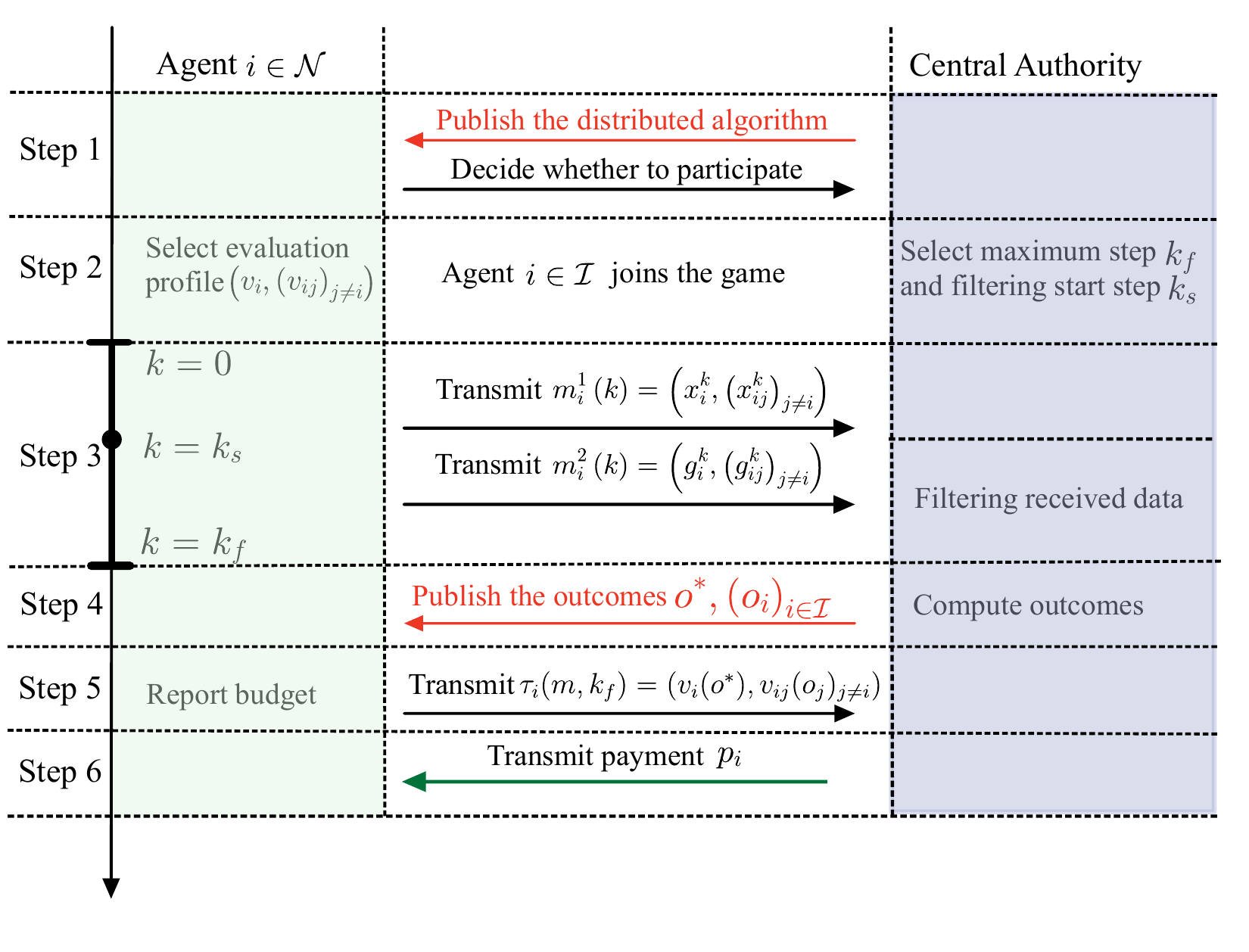}}
\caption{Steps of the proposed DeVCG-G mechanism}
\label{fig3-2}
\end{figure}

Figure \ref{fig3} shows an example of the information flow among agents and the message flow between the agent and the central unit. In cases that inter-agent data \(z^k_i\) and agent-central messages \(m^1_i\) or \(m^2_i\) contain overlapping information, the overlapping portions must be identical. Figure \ref{fig3-1} shows the computation structure of the central unit. The proposed mechanism is shown in Mechanism 2 and the steps are summarised in Fig. \ref{fig3-2}.

\textbf{DeVCG-G Game:} The DeVCG-G Mechanism induces
a DeVCG-G Game where each agent \(i\in\mathcal{N}\) selects \(v_{i}\in\mathcal{V}\) for the social sequence and \(v_{ij}\in\mathcal{V}\) for sequence \(j\). In particular, quitting agent \(i\)'s strategies is \(v_{i}=v_{ij}=\emptyset, \forall j\neq i,i\in\mathcal{N}\setminus\mathcal{I}, \). Each agent \(i\in\mathcal{I}\) simultaneously computes \(x^k_i\) and \(x^k_{ij}, \forall j\neq i\) at step \(k\). The sequences \(\{x^k_i\}_{0\leq k\leq k_f}\) and \(\{x^k_{ij}\}_{0\leq k\leq k_f}\) are generated by the specified distributed algorithm (\ref{4}):
\begin{subequations}
\label{27}\begin{align}
    x^k_i&=[\sigma(k,x_0,v)]_i,\quad v=(v_i,v_{-i})\\
    x^k_{ij}&=[\sigma(k,x_0,\text{v}_{-j})]_i,\quad \text{v}_{-j}=(v_{pj})_{p\neq j}
\end{align}
\end{subequations}
The auxiliary gradient sequences are calculated as: 
\begin{subequations}
\begin{align}
g^k_i&\in\partial v_i(x^k_i)\\
g^k_{ij}&\in\partial v_{ij}(x^k_{ij})
\end{align} 
\end{subequations}
and the filter starts filtering the gradient data at step \(k_s<k_f\).
Denote:
\begin{subequations}
\begin{align}
\nu_{i}&=\bigl(v_{i}, ({v}_{ij})_{j\neq i}\bigr)\label{niu i}\\
\nu_{-i}&=\bigl(v_{-i}, (\text{v}_{-j})_{-i}\bigr)_{j\neq i}\label{niu -i}
\end{align}
\end{subequations}
where \((\text{v}_{-j})_{-i}=(v_{pj})_{p\in\mathcal{N}\setminus\{i,j\}}\in\mathcal{V}^{N-2}\)

We use the notation \(m_i(\nu,k)\) to represent agent \(i\)'s message at step \(k\) when the overall evaluation profile is \(v\) in the social sequence, \(\text{v}_{-j}\) in sequence \(j\). 
The outcomes \(o^*(m,k_f)\) and \(o_i(m_{-i},k_f)\) are selected by the central authority as the component-wise median of \(\{x^{k_f}_i\}_{i\in\mathcal{I}}\) and \(\{x^{k_f}_{ji}\}_{j\neq i}\), respectivelly. The decided outcomes are sent back to each agent. 
 The notations \(o^*(m,k_f)\) and \(o_i(m_{-i},k_f)\) characterise the dependence of outcomes on the message and the maximum iteration step.

The agent \(i\in\mathcal{I}\) decides the budget proposal as:
\begin{align}
    \tau_{i}(m,k_f)=(v_i(o^*),v_{ij}(o_j)_{j\neq i})\label{tau}
\end{align}
In particular, to clarify the notation as used in (\ref{18}), we denote \((\tau_{i}(m,k_f))_0=v_i(o^*)\) and \((\tau_{i}(m,k_f))_j=v_{ij}(o_j)\).
Upon receiving the budget proposal, the central authority determines payment \(p_i(m,k_f)\) for each agent \(i\in\mathcal{I}\) and \(p_i=0\) for \(i\in\mathcal{N\setminus}\mathcal{I}\).

In this game, strategic and malicious agents aim at selecting an evaluation function profile \((v_i,v_{ij})_{j\neq i}\) to maximise their own payoff \(u_i(o^*,p_i)=-f_i(o^*)-p_i\) and minimise others' payoff if his own payoff is not influenced. \hfill \(\square\)
\begin{algorithm}
\caption{DeVCG-G Mechanism}
\begin{algorithmic}[1]
    \State \textbf{Initialisation}
    
     The central authority publishes the distributed algorithm (whose dynamic is (\ref{4})), the termination tolerance \(\varepsilon\) and the maximum iteration step \(k_f\). It is guaranteed by the central authority that the tolerance is satisfied at a step \(k\leq k_f\).
    \State \textbf{The message space \(\mathcal{M}=\Pi_i \mathcal{M}_i\), \(\mathcal{M}_i\subset \mathcal{X}^{nI(k_f+1)}\times\mathbb{R}^{nI(k_f+1)}\times\mathbb{R}^{I}\)} 
    
    The message \(m_i\in\mathcal{M}_i\) for each agent contains two parts \(m_i:=\hat{m}_i\times \{\tau_{i}(\hat{m}^1,k_f)\}\). The first part \(\hat{m}_i\) including \(m^1_i(k)\in\mathcal{X}^{nI(k+1)}\) and \(m^2_i(k)\in\mathbb{R}^{nI(k+1)}\) are dynamically submitted and the second part is submitted after the outcomes have be determined.
      \begin{subequations}
      \label{20_2}
      \begin{align}
      m_i:=&\hat{m}_i\times \{\tau_{i}(\hat{m},k_f)\}\\
      \hat{m}_i:=&\Pi_k\bigl(m^1_i(k)\times m^2_i(k)\bigr)\\
     m^1_i(k):=&\bigl\{\bigl(x^{k}_i, (x^{k}_{ij})_{j\neq i}\bigr)\bigr\}\\
     m^2_i(k):=&\bigl\{ \bigl(g^{k}_{i}, (g^{k}_{ij})_{j\neq i}\bigr)\bigr\}
     \end{align}
     \end{subequations}
     where \((x^k_{i},(x^k_{ij})_{j\neq i})\) are as in (\ref{27}) representing agent \(i\)'s states in the social sequence and in sequence \(j\) at step \(k\), \((g^k_{i},(g^k_{ij})_{j\neq i})\) are auxiliary data  
     and \(\tau_{i}\) is agent \(i\)'s budget proposal which depends on \(\hat{m}=(\hat{m}_i,\hat{m}_{-i})\) and the termination step \(k_f\). 

    \State \textbf{Social \(\&\) sequences outcome \(o^*,o_i\in\mathbb{R}^n,i\in\mathcal{I}\)} 
   The social outcome \(o^*(m,k_f)\) is taken as the component-wise median of \(\{x^{k_f}_i\}_{i\in\mathcal{I}}\) and sequence \(i\)'s outcome \(o_{i}(m_{-i},k_f)\) is taken as the component-wise median of \(\{x^{k_f}_{ji}\}_{j\neq i}\):
    \begin{subequations}\label{outcome selection}
   \begin{align}
       o^*(m,k_f)&=Median(x^{k_f}_i)_{i\in\mathcal{I}}\\
       o_i(m_{-i},{k_f})&=Median(x^{k_f}_{ji})_{j\neq i}
   \end{align}
   \end{subequations}
   
    \State \textbf{Payment \(p_i\in\mathbb{R},i\in\mathcal{N}\)} 

     If all agents quit the game, the payment for each agent \(i\in\mathcal{N}\) is \(\bar{p}\), where \(\bar{p}<+\infty\) is positive and sufficiently large.
     
     Otherwise, the payment for each agent \(i\in\mathcal{N}\setminus\mathcal{I}\) is \(p_i=0\).
    and the payment for each agent \(i\in\mathcal{I}\) is calculated as:
     \begin{align}         
         p_i(m,k_f):=&\sum_{j\neq i}(\tau_{j}(m,k_f))_0-\sum_{j\neq i}(\tau_{j}(m,k_f))_i\notag\\
         &+\pi_i(m,k_f)\label{18}
     \end{align}
    where 
     \begin{subequations}
     \label{25}\begin{align}
        &\pi_i(m,k_f) = 
            \begin{cases}
            0,\qquad\qquad\qquad\qquad\,\text{if }e_i=0\\
            k_f\cdot e_i(m,k_f)+1,\,\quad\text{otherwise}
            \end{cases}\\
&e_i(m,k_f)=\sum_{t=Ik_s}^{I(k_f+1)-1}\|\Tilde{\xi}_{i}^t-\xi_i^t\|^2-\sum_{j\neq i}\min\bigl\{0,(\tau_{i})_j\notag\\
&-({\tau}_{i})_0-(g^{k_f}_{i})^T(o_j-o^*)\bigr\}\label{18b}
     \end{align}
    with the first term in (\ref{18b}) be calculated as in (\ref{17c}). 

\end{subequations}
\end{algorithmic}
\end{algorithm}

We emphasise that each agent's final step state-budget pair \(m_i(\nu,k_f)\) has the following form:
\begin{align}
    m_i(\nu,k_f)=(x^{k_f}_i,(x^{k_f}_{ij})_{j\neq i})\times(v_i(o^*), v_{ij}(o_j)_{j\neq i})\label{34}
\end{align}

Given a collection of \(\nu_{-i}\in\mathcal{V}^{N-1}\times(\mathcal{V}^{N-2})^{N-1}\) (as defined in (\ref{niu -i})) for other agents \(j\neq i\) and the total iteration step \(k_f\). The space of all possible final step state-budget pair from agent \(i\)'s perspective is:
\begin{align}
    \textbf{M}(\nu_{-i},k_f)&=\bigcup_{\nu_i\in\mathcal{V}^N}\Bigl\{\bigl(m_i(\nu_i,\nu_{-i},k_f),m_{j}(\nu_i,\nu_{-i},k_f)_{ j\neq i}\bigr) \notag\\
    &\text{ as in } (\ref{34}) :\,x^{k_f}_i,x^{k_f}_{ij},x^{k_f}_j,x^{k_f}_{jp}  \text{ are as in } (\ref{27})\notag\\
    & \text{ and }o^*,o_i,o_j \text{ are as in } (\ref{outcome selection}),\forall j,p\neq i\Bigr\}
\end{align}

Similar to the DeVCG mechanism, we introduce the following definition for the DeVCG-G game induced by the DeVCG-G mechanism.

\textbf{Definition 9.} An \(\varepsilon\)-DSE of the DeVCG-G game for any given \(\varepsilon\in(0,1)\) is defined as a final step state-budget profile \(m^*=(m^*_i,m^*_{-i})\in\cap_i{\textbf{M}}(\nu^*_{-i},k_f)\) corresponding to an evaluation profile \((\nu^*_i,\nu^*_{-i})\) with \(\nu^*_i=(v^*_i,v^*_{ij})_{j\neq i}\). For all \(m=(m_i,m_{-i})\in\cup_i{\textbf{M}}(\nu^*_{-i},k_f)\) and all \(i\in\mathcal{N}\), there exists a finite step \(k\) and the following holds for any \(k_f\geq k\):
\begin{align}
    &u_i(o^*(m^*{k_f}),p_i(m^*,{k_f}))+\varepsilon\\
    &\geq u_i(o^*(m,{k_f}),p_i(m,{k_f}))\notag
\end{align}
where \(o^*(m^*,k_f)=Median(x^{k_f}_i)_{i\in\mathcal{I}}\) and \(o^*(m,k_f)=Median(\hat{x}^{k_f}_{i})_{i\in\mathcal{I}}\). With an appropriate collection of initial states \(x_0\), the collection of \(x^{k_f}_i\) and \(\hat{x}^{k_f}_{i}\) are calculated as:
\begin{subequations}
\begin{align}
    x_{k_f}&=\sigma({k_f},x_0,v^*)\\    \hat{x}_{k_f}&=\sigma({k_f},x_0,v_i,v^*_{-i}) 
\end{align}
\end{subequations}
Also, \(p_i(m^*,k_f)\) and \(p_i(m,{k_f})\) are computed as \eqref{18} and \eqref{tau} based on \(\nu^*\) and \((\nu_i,\nu_{-i}^*)\), respectively.

\section{Properties analysis}
In this section, we investigate the properties of the DeVCG-G mechanism with agents adopting two different manipulation strategies. We establish that
the designed mechanism possesses the desired properties. Also, the induced game has \(\varepsilon\)-dominant strategy equilibria, which can be reached by implementing the distributed algorithm.
\subsection{Sequence-independent manipulation strategies}
Notice that under TISI manipulation strategies, \(e_i(m,k_f)=0\) holds for all agent \(i\in\mathcal{I}\).
Therefore, the DeVCG-G game is equivalent to the DeVCG game. Directly from Theorem 1, we obtain the following theorem.

\textbf{Theorem 2} Let Assumptions 1 and 2 hold, and let all agents use TISI manipulation strategies. Given any tolerance \(\varepsilon\in(0,1)\), there exists \(k\) such that the DeVCG-G mechanism \(\Gamma_k\) is \(\varepsilon\)-IC and the sequence \(\{\Gamma_{k_f}\}_{k_f\geq k}\) is asymptotic efficient.
\subsection{Sequence-dependent manipulation strategies}
Next, we consider agents using the TISD manipulation strategy. Notice that the TISI manipulation strategy is a special case of the TISD manipulation strategy. We first introduce a preparatory lemma 2 (Lemma 3.3 in \cite{gonzalez2023monotone}, Theorem B in \cite{rockafellar1970maximal}).

\textbf{Lemma 2.} Let \(v_1, v_2:\mathbb{R}^n\to(-\infty,+\infty]\) be proper lower semi-continuous convex functions such that their subdifferentials overlap everywhere in an open convex set \(\mathcal{X}\subset \text{int(dom}(v_1))\cap\text{int(dom}(v_2))\):
\begin{align}
    \partial v_1(x)\cap\partial v_2(x)\neq \emptyset
\end{align} 
Then \(v_1(x) = v_2(x)+constant, \forall x\in\mathcal{X}\).

\textbf{Lemma 3.} For any given \(\varepsilon\in (0,1)\) and \(k_f\) sufficiently large, at any \(\varepsilon\)-DSE of the DeVCG-G game \({m}^*=({m}^*_i,{m}^*_{-i})\in\textbf{M}(\nu^*_{-i},k_f)\), we have:
\begin{align}
    e_i({m}^*,k_f)=0
\end{align}
holds for all agent \(i\in\mathcal{N}\).

\textbf{Proof.} We prove this lemma by contradiction. Suppose there exists an \(\varepsilon\)-DSE \(\hat{m}=(\hat{m}_i,\hat{m}_{-i})\in \textbf{M}(\nu^*_{-i},k_f)\) induced by agent \(i\)'s evaluation profile \(\hat{\nu}_i\) leading to \(e_i(\hat{m},k_f)>0\) for agent \(i\). According to (\ref{25}) and lemma 2, his evaluation function profile is \(\hat{\nu}_i=(\hat{v}_i,\hat{v}_{ij})_{j\neq i}\) with the condition that \(\exists x\in\mathcal{X}\), s.t. \(\partial \hat{v}_i(x)\cap \partial \hat{v}_{ij}(x)=\emptyset\) or \(\hat{v}_{ij}(\cdot)=\hat{v}_i(\cdot)+c_{ij}\), where \(c_{ij}<0\) is a constant and fulfils the following inequality:
\begin{align}
    \hat{v}_{ij}(o_j)=\hat{v}_i(o_j)+c_{ij}<\hat{v}_i(o^*)+\nabla \hat{v}_{i}(o^*)^T(o_j-o^*)
\end{align}
where \(o^*\) and \(o_j\) are the outcomes in social sequence and sequence \(j\), respectively.
Now, consider a deviation of agent \(i\) from \(\hat{\nu}_i=(\hat{v}_i,\hat{v}_{ij})_{j\neq i}\) to \(\nu_i=(\hat{v}_i,\hat{v}_i,\cdots,\hat{v}_i)\). Let \((m_i,m_{-i})\in \textbf{M}(\nu^*_{-i},k_f)\) be obtained by using evaluation profile \((\hat{v}_i,\hat{v}_i,\cdots,\hat{v}_i)\) and \(\nu^*_{-i}\). Since \(\hat{m}\) is an \(\varepsilon\)-DSE, we have:
\begin{align}
    &u_i(o^*(\hat{m},k_f),p_i(\hat{m},k_f))+\varepsilon\\
    &\geq u_i(o^*({m},k_f),p_i({m},k_f)) 
\end{align}
which leads to the following inequality:
\begin{align}
    &f_i(o^*(\hat{m},,k_f))+p_i(\hat{m},,k_f)\notag\\
    &\leq f_i(o^*({m},k_f))+p_i({m},k_f)+\varepsilon 
\end{align}
Noticing that the social outcome and sequence \(i\)'s outcome remain unchanged \(o^*(\hat{m},k_f)=o^*({m},k_f)\), the following holds \(p_i(\hat{m},k_f)-p_i({m},k_f)=\pi_i(\hat{m},k_f)\). Hence, we obtain:
\begin{align}
    \pi_i(\hat{m},k_f)\leq \varepsilon \label{32}
\end{align}
By construction if \(e_i>0\), we have \(\pi_i>1\). Because \(\varepsilon<1\), inequality (\ref{32}) indicates \(\pi_i(\hat{m},k_f)=0\)
and therefore \(e_i(\hat{m},k_f)= 0\). This contradicts the supposition that \(e_i(\hat{m},k_f)>0\).\hfill \(\square\)

An immediate consequence of Lemma 3 is that the maliciousness of agents is bounded. 

Assume that agent \(i\)'s evaluation profile \(\nu_i=(v_i,v_{ij})\) contribute to an \(\varepsilon\)-DSE \(m^*\). Since \(e_i(m^*,k_f)=0\), according to the first term in (\ref{18b}), there exists a convex function \(v\) such that the following holds for all \(k\in[0,k_f]\) and all \(i,j\in\mathcal{I}\):
\begin{align}
    &\partial v_i(x^k_i)\cap\partial v(x^k_i)\neq \emptyset\\
    &\partial v_{ij}(x^{k}_{ij})\cap\partial v(x^k_{ij})\neq \emptyset
\end{align}
The realisations of sequences \(\{x^k_i\}\) and \(\{x^k_{ij}\}\) depend on the selected distributed algorithm, the evaluation profile, the collection of initial states and the overall iteration steps -- all of which are unpredictable by an individual agent. Selecting functions \(v_i(\cdot),v_{ij}(\cdot)\) such that \(\exists x\in\mathcal{X} \) with \(\partial v_i(x)\cap\partial v_{ij}(x)=\emptyset\) carries uncertain additional cost induced by the gradient filter. We make the following assumption.

\textbf{Assumption 3 (Conservative Adversarial Behaviour).}  We assume all agents avoid taking strategies that could activate the gradient filter. 

The assumption states that agents prioritise maximising their payoff over being malicious to others. Under Assumption 3, agents' evaluation functions fulfil \(\partial v_i(x)\cap\partial v_{ij}(x)\neq\emptyset,\forall x\in\mathcal{X}\). According to Lemma 2, we conclude that \(v_{ij}(\cdot)=v_i(\cdot)+c_{ij}\). The non-positive constants \(c_{ij},j\neq i\) represent the maliciousness of agent \(i\) in sequence \(j\). Let \(o^*\) be the social outcome and \(o_j\) be sequence \(j\)'s outcome, the following bounds hold by the second term in (\ref{18b}):
\begin{align}
    &v_i(o^*)+\nabla v_i(o^*)^T(o_j-o^*)-v_i(o_{j})\leq c_{ij}\leq 0
\end{align}

\textbf{Lemma 4.} Let Assumptions 1 -- 3 hold, and let all agents join the game. Given any tolerance \(\varepsilon\in(0,1)\), there exists a finite \(k\) such that for all \(k_f\geq k\), any \(\varepsilon\)-DSE \(m^*=(m_i^*,m^*_{-i})\) has the following form:
\begin{align}
    m_i^* = \bigl(x_i^{k_f},(x_{ij}^{k_f})_{j\neq i}\bigr)\times \Bigl(f_i(o^*),\bigl(f_{i}(o_{j})+c_{ij}\bigr)_{j\neq i}\Bigr)\label{36}
\end{align}
where given a collection of proper initial states \(x_0\):
\begin{subequations}
\begin{align}
  x_i^{k_f}&=[\sigma\bigl(k_f,x_0,f_i(\cdot),f_{-i}(\cdot)\bigr)]_i \label{51a}\\
    x_{ij}^{k_f}&=[\sigma\bigl(k_f,[x_0]_{-j},f_i(\cdot)+c_{ij},f_{p}(\cdot)+c_{pj}\bigr)]_i, \forall p\neq i,j
\end{align}
with \(o^*=Median(x^{k_f}_i)_{i\in\mathcal{N}},o_j=Median(x^{k_f}_{ij})_{j\neq i}\).
\end{subequations}
Any selection of constants \(c_{ij}\) fulfils:
\begin{align}
   f_i(o^*)+\nabla f_i(o^*)^T(o_j-o^*)-f_i(o_j)\leq c_{ij}\leq 0 \label{38}
\end{align}
Among the set of \(\varepsilon\)-DSE, selecting \(c_{ij}=0,\forall i,j\) lead to a Pareto \(\varepsilon\)-DSE.

\textbf{Proof.} According to Lemma 3, at any \(\varepsilon\)-DSE \(m^*=(m_i^*,m^*_{-i})\in\textbf{M}(\nu^*_{-i},k_f)\) with \(\varepsilon\in(0,1)\), \(e_i(m^*,k_f)=0\). This indicates that the evaluation profile for agent \(i\) has the form: \(\nu_i=(v^*_i(\cdot),v^*_{i}(\cdot)+\bar{c}_{ij})\) with
\begin{align}
    &v^*_i(o^*)+\nabla v^*_i(o^*)^T(o_j-o^*)-v^*_i(o_{j})\leq \bar{c}_{ij}\leq 0
\end{align}
and the message has the following form:
\begin{align}
    m_i^* = \bigl(x_i^{k_f},(x_{ij}^{k_f})_{j\neq i}\bigr)\times \Bigl(v^*_i(o^*),\bigl(v^*_{i}(o_{j})+\bar{c}_{ij}\bigr)_{j\neq i}\Bigr)
\end{align}
Now, suppose \(v^*_i\neq f_i\) and let the message profile \(\hat{m}^*\in\textbf{M}(\nu^*_{-i},k_f)\) induced by \((f_i(\cdot),f(\cdot)+{c}_{ij})\) be as in \eqref{36} and \eqref{38}, by definition of  \(\varepsilon\)-DSE:
\begin{align}
    &u_i\bigl(o^*(m^*,k_f), p_i(m^*,k_f)\bigr)-u_i(o^*(\hat{m}^*,k_f),p_i(\hat{m}^*,k_f))\notag\\
    &=f_i(o^*(\hat{m}^*,k_f))+\sum_{j\neq i} v^*_{j}(o^*(\hat{m}^*,k_f))\notag\\
    &-\bigl(f_i(o^*(m^*,k_f))+\sum_{j\neq i} v^*_{j}(o^*(m^*,k_f))\bigr)\notag\\
    &\geq \varepsilon 
\end{align}
which leads to the following inequality for all \(k_f\) sufficiently large:
\begin{align}
    &f_i(o^*(\hat{m}^*,k_f))+\sum_{j\neq i} v^*_{j}(o^*(\hat{m}^*,k_f))\label{55_1}\\
    &\geq f_i(o^*(m^*,k_f))+\sum_{j\neq i} v^*_{j}(o^*(m^*,k_f))+\varepsilon\notag
\end{align}
However, according to \eqref{51a}, we see 
\begin{align}
    &\lim_{k_f\to+\infty}f_i(o^*(\hat{m}^*,k_f))+\sum_{j\neq i} v^*_{j}(o^*(\hat{m}^*,k_f))\\
    &=\min_x\{f_i(x)+\sum_{j\neq i}v^*_j(x)\}\notag
\end{align}
contradicts \eqref{55_1} and hence contradicting the supposition that \(v^*_i\neq f_i\). 

The payoff of agent \(i\) at \(\varepsilon\)-DSE \(m^*\) is:
\begin{align}
    &u_i\bigl(o^*({m}^*,k_f),p({m}^*,k_f)\bigr)\notag\\
    &=-\sum_{i\in\mathcal{N}}f_i(o^*({m}^*,k_f))+\sum_{j\neq i}f_j(o_i({m}^*_{-i},k_f))+\sum_{j\neq i}c_{ji}\notag
\end{align}
Among these \(\varepsilon\)-DSE, since \(c_{ji}\) is non-positive, we see that when all agents set \(c_{ji}=0\), they obtain the highest payoff.\hfill\(\square\)

At \(\varepsilon\)-DSE \(m^*\), Agent \(i\)'s maximum loss caused by others' maliciousness is: 
\begin{align}
    &\bigl|\sum_{j\neq i}c_{ji}\bigr|=\sum_{j\neq i}f_{j}(o_i(m^*_{-i},k_f))-\sum_{j\neq i}f_{j}(o^*(m^*,k_f))\label{41}\\
    &-\sum_{j\neq i}\nabla f_j(o^*(m^*,k_f))^T(o_i(m^*_{-i},k_f)-o^*(m^*,k_f))\notag\\
    &\leq \sum_{j\in\mathcal{N}}f_{j}(o_i(m^*_{-i},k_f))-\bigl[\sum_{j\in\mathcal{N}}f_{j}(o^*(m^*,k_f))\notag\\
    &+\nabla\sum_{j\in\mathcal{N}} f_j(o^*(m^*,k_f))^T(o_i(m^*_{-i},k_f)-o^*(m^*,k_f))\bigr]\notag\\
    &\leq \sum_{j\in\mathcal{N}}f_{j}(o_i(m^*_{-i},k_f))-\sum_{j\in\mathcal{N}}f_{j}(o^*(m^*,k_f))\notag\\
    &+\Bigl\|\nabla\sum_{j\in\mathcal{N}} f_j(o^*(m^*,k_f))\Bigr\|\bigl\|o_i(m^*_{-i},k_f)-o^*(m^*,k_f)\bigr\|\notag
\end{align}
Because the maliciousness \(c_{ij},\forall i,j\) in each sequence is decided before the game begins, the above relationship holds for all possible \(k_f\).

Let \(x^*\) be the unique minimiser of cost function \(\sum_{i\in\mathcal{N}}f_i(x)\), we have \(0\in\partial \sum_{i\in\mathcal{N}}f_i(x^*)
\) holds. Hence:
\begin{align}
    &\liminf_{k_f\to+\infty} \Bigl\|\nabla\sum_{j\in\mathcal{N}} f_j(o^*(m^*,k_f))\Bigr\|\bigl\|o_i(m^*_{-i},k_f)-o^*(m^*,k_f)\bigr\|\notag\\
    &=0\label{42}
\end{align}
For any \(\varepsilon>0\), there exists a finite \(k\) such that:
\begin{align}
    &\inf_{k_f\geq k}\Bigl\{\Bigl\|\nabla\sum_{j\in\mathcal{N}} f_j(o^*(m^*,k_f))\Bigr\|\bigl\|o_i(m^*_{-i},k_f)-o^*(m^*,k_f)\bigr\|\Bigr\}\notag\\
    &\leq \varepsilon
\end{align}
We then conclude that:
\begin{align}
    \bigl|\sum_{j\neq i}c_{ji}\bigr|\leq \sum_{j\in\mathcal{N}}f_{j}(o_i(m^*_{-i},k_f))-\sum_{j\in\mathcal{N}}f_{j}(o^*(m^*,k_f))+\varepsilon\label{43}
\end{align}
This relationship remains true if there is only a subset \(\mathcal{I}\subseteq\mathcal{N},\mathcal{I}\neq\emptyset\) of agents joining the game.

\textbf{Lemma 5.} Let Assumptions 1 -- 3 hold, for any given \(\varepsilon\in(0,1)\), the DeVCG-G mechanism incentivises agents to join the game.

\textbf{Proof.} Let us now compare the case where a set of agents \(i\in\mathcal{N}\setminus\mathcal{I}\) do not want to join the game, which is \(v_i=\emptyset,v_{ij}=\emptyset,\forall j\neq i\), with the case where an agent \(i\in\mathcal{N}\setminus\mathcal{I}\) joins the game. We use the notation \((\emptyset,\hat{m}^*_{-i})\in\textbf{M}(\nu^*_{-i},k_f)\) to denote the last step state-budget profile when the DeVCG-G game with agents \(j\in\mathcal{I}\) achieve an \(\varepsilon\)-DSE as in (\ref{36}). Let \({m}^*=(m^*_i,m^*_{-i})\in\textbf{M}(\nu^*_{-i},k_f)\) denote an \(\varepsilon\)-DSE with agents \(\{i\}\cup\mathcal{I}\) joining the game. Then, the following holds:
\begin{align}
     &u_i(o^*(\emptyset,\hat{m}^*_{-i},k_f),0)-u_i(o^*({m}^*,k_f),p_i({m}^*,k_f))\\
     &=\bigl[f_i(o^*({m}^*,k_f))+\sum_{j\in\mathcal{I}}f_j(o^*({m}^*,k_f))\notag\\
     &-\sum_{j\in\mathcal{I}}f_j(o_i({m}_{-i}^*,k_f))-\sum_{j\in\mathcal{I}}c_{ji}\bigr]-f_i(o^*(\emptyset,\hat{m}_{-i}^*,k_f))\notag\\
     &\stackrel{(a)}{=} \sum_{j\in\mathcal{I}\cup\{i\}}f_j(o^*({m}^*,k_f))-\sum_{j\in\mathcal{I}\cup\{i\}}f_j(o_i({m}_{-i}^*,k_f))\notag\\
     &+\bigl|\sum_{j\in\mathcal{I}}c_{ji}\bigr|\notag\\
     &\stackrel{(b)}{\leq}\varepsilon\notag
\end{align}
where (a) holds because \(o_i({m}_{-i}^*,k_f)=o^*(\emptyset,\hat{m}_{-i}^*,k_f)\) and (b) holds according to \eqref{43} by changing \(\mathcal{N}\) to \(\mathcal{I}\cup\{i\}\).  
Therefore, the DeVCG-G mechanism incentivises agents to join the game.\hfill\(\square\)

As a result of Lemma 4 and 5, we present the following theorem.

\textbf{Theorem 3.} Let Assumptions 1 -- 3 hold and let all agents use TISD manipulation strategies. Given any tolerance \(\varepsilon\in(0,1)\), there exists \(k\) such that the DeVCG-G mechanism \(\Gamma_k\) is \(\varepsilon\)-IC and the sequence \(\{\Gamma_k\}\) is asymptotic efficient. 

\section{Illustrative example}
\begin{figure}
\centerline{\includegraphics[width=1\columnwidth]{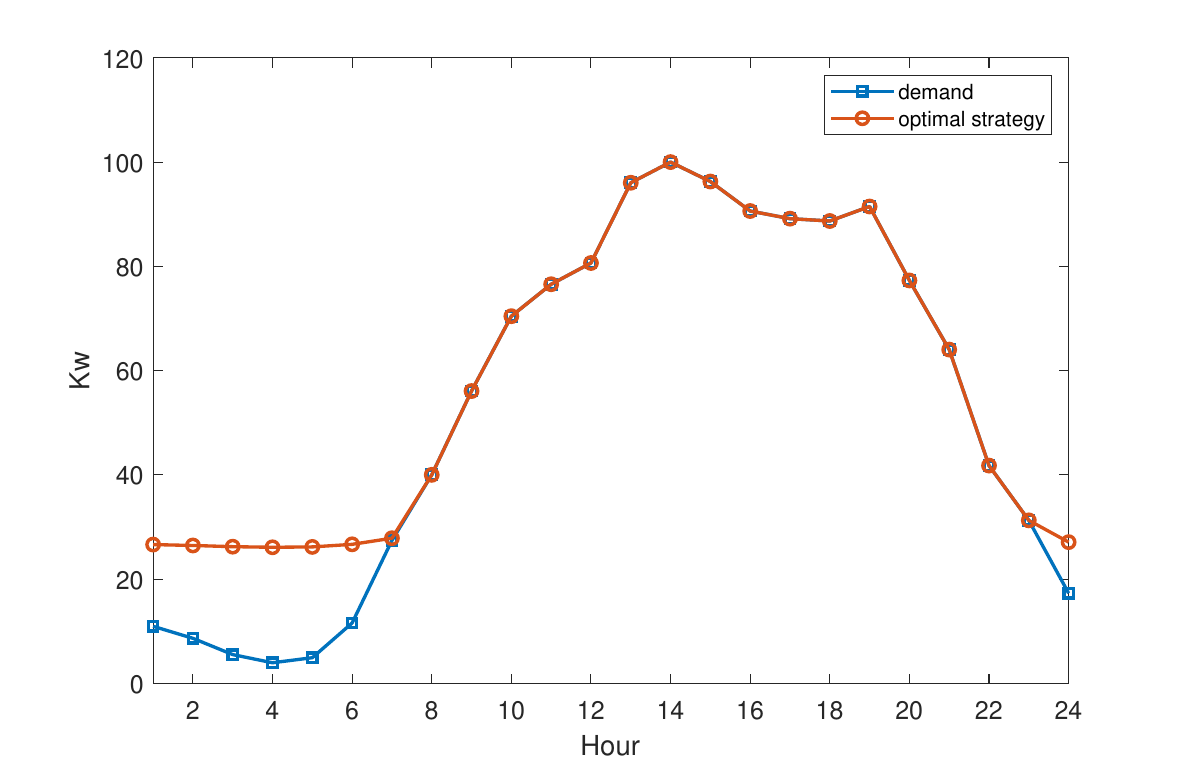}}
\caption{Background demand and the optimal charging strategies.}
\label{fig6}
\end{figure}

We present an illustrative example of an electric vehicle (EV) charging coordination problem. The problem and optimisation model is formulated by \cite{zou2016efficient}. Consider a population of \(N\) EVs charging over a common charging horizon \(\mathcal{T}=\{t_0,t_0+\Delta T,\dots,t_0+(n-1)\Delta T\}\). We consider the time horizon from 1:00 to 24:00 on one day and set \(n=24\) with a time interval of \(\Delta T=1\)h. Also denote \(t_T\) as \(t_T=t_0+23\Delta T\). For each EV, \(i\in \mathcal{N}=\{1,2,\dots,N\}\), the energy delivered over the \(t\)-th time interval is \(x_{it}\) and the cost function of each EV \(i\in\mathcal{N}\) is:
\begin{align}
    f_i(\bm{x}_i) &= \sum_{t\in\mathcal{T}} w_i(x_{it}) + \alpha_i (\sum_{t\in\mathcal{T}}x_{it}-\Gamma_i)^2\notag\\
&+\frac{\beta}{N}\sum_{t\in\mathcal{T}}(D_t+\sum_{i\in\mathcal{N}}x_{it})^{2}+200+\gamma_i
\end{align}
where \(D_t\) is the aggregate demand at time \(t\), \(\alpha_i\), \(\gamma_i\) are locally known parameters and \(\beta\) is a commonly known parameter. 
The function \(w_i(\cdot)\) represents the battery degradation cost and \(\Gamma_i\) is the maximum energy that can be delivered:
\begin{subequations}
\begin{align}
    w_i(x_{it})&=0.002x_{it}^2\\
    \Gamma_i&=\Theta(\Bar{s}-s_{i,0})
\end{align}
\end{subequations}
where \(\Theta\) is a common battery capacity, \(\bar{s}\) is a common maximum battery state of charge (SoC)
value and \(s_{i,0}\) is the initial SoC value for \(i\)-th EV. An admissible charging strategy \(\bm{x}_i=[x_{it_0},\dots,x_{it_T}]'\) satisfies the following constraints:
\begin{align}
    \sum_{t\in \mathcal{T}} x_{it}\leq \Gamma_i \notag
\end{align}
For a collection of admissible charging strategy \(\bm{x}=[\bm{x}_1,\cdots,\bm{x}_N]'\), the social minimisation problem we are going to solve is:
 \begin{align}
     \min_{\bm{x}}&\,\sum_{i=1}^N f_i({\bm{x}})\\
     {\rm s.t.}&\,\sum_{t\in \mathcal{T}} x_{it}\leq \Gamma_i \notag \notag
 \end{align}
 The demand profile \(\bm{D}=(D_t),t\in \mathcal{T}\) is presented in Fig. \ref{fig6}. 
 
In this example, we consider \(N=4, \beta = 0.005\) and the truethful parameter \(\alpha_i\) and the initial SoC values \(s_{i,0},{i\in\mathcal{N}}\) for each agent is \(\alpha=[10, 4, 8, 7]\) and \(s_{0}=[0.1, 0.15, 0.23, 0.14]\), respectively. Also, set  \(\Theta=30{\rm kWh}\) and \(\Bar{s}=0.9\). The decision space narrows down from the evaluation function space to a parameter space. An agent \(i\) can be self-interested by deviating \(\alpha_i\) from its true value in all sequences and be malicious by decreasing \(\gamma_i\) in sequences \(j\neq i\).
 

\subsection{TISI manipulation strategies}
We consider agents are using TISI manipulation strategies. Each agent \(i\in\mathcal{N}\) choose \(\alpha_i\) before join the game. The following figure Fig. \ref{fig7} confirms that all agents obtain their maximum payoff when using true \(\alpha_i\). 
\begin{figure}
\centering
    \begin{subfigure}[b]{0.24\textwidth}      \centering\includegraphics[width=1\textwidth]{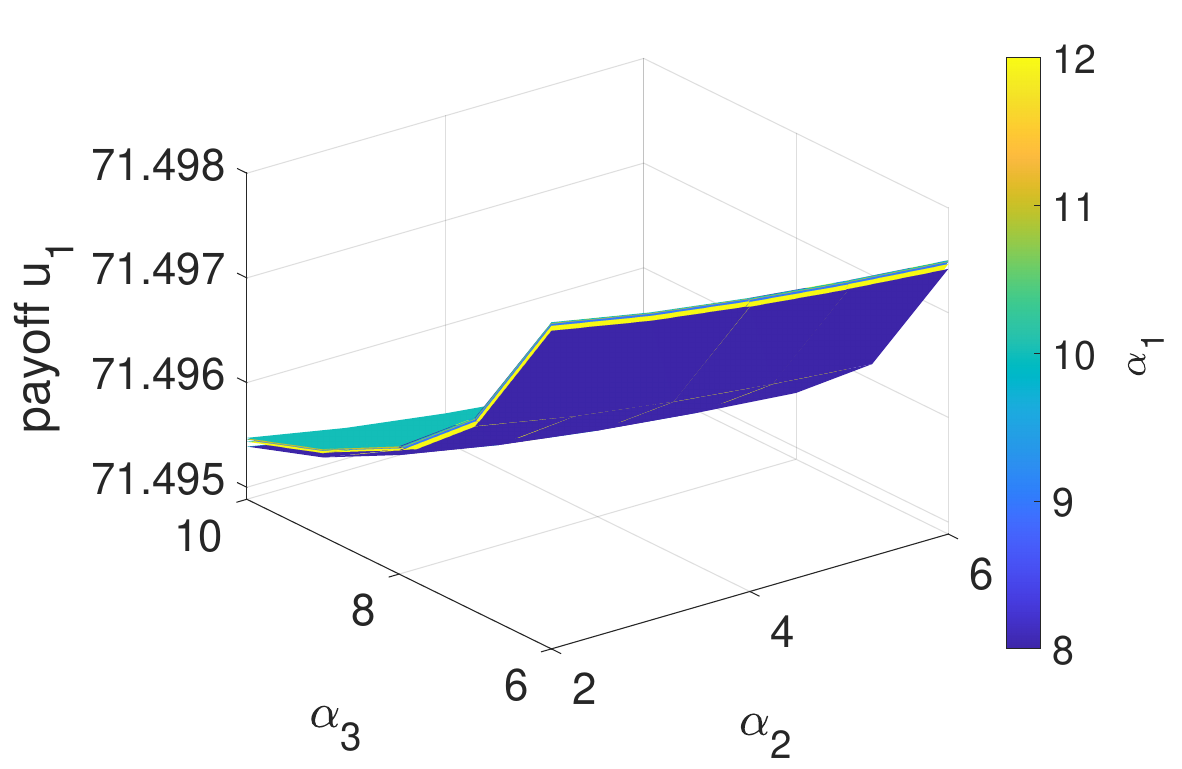}
        \label{Fig.sub.11}
        \caption{}
    \end{subfigure}
        \begin{subfigure}[b]{0.24\textwidth}
        \centering
\includegraphics[width=1\textwidth]{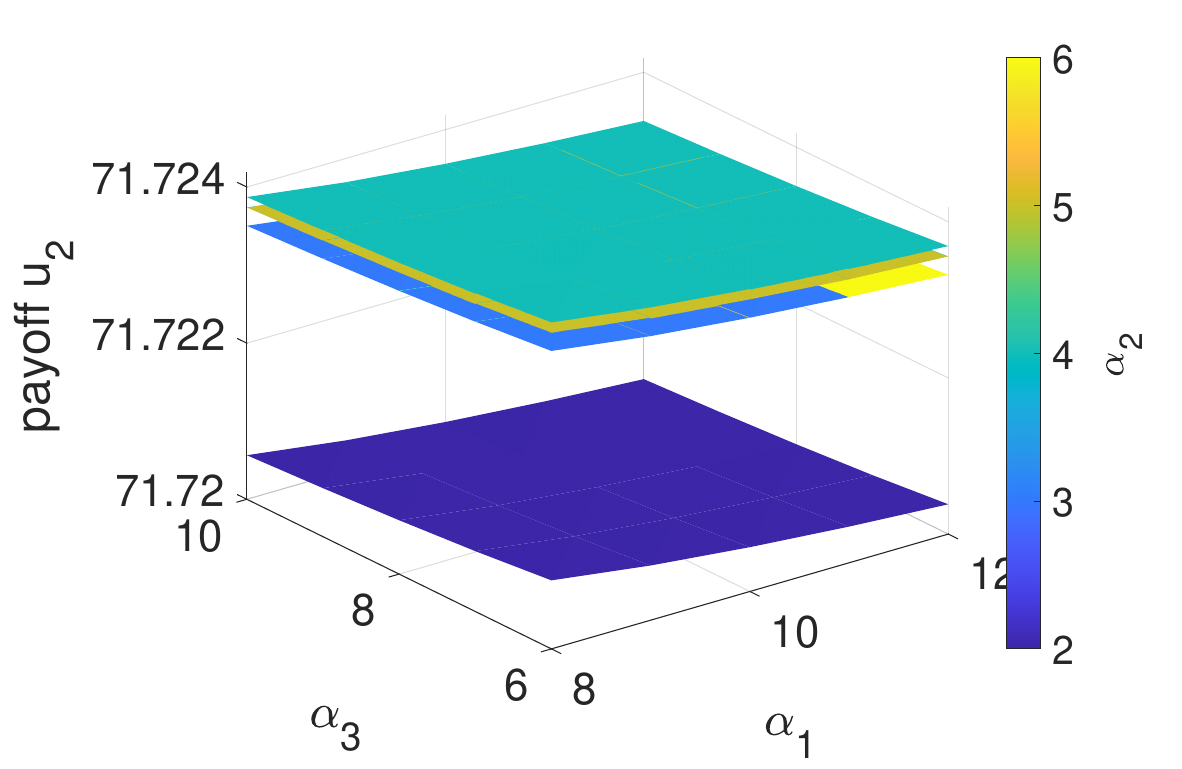}
        \label{Fig.sub.12}
         \caption{}
    \end{subfigure}
    \begin{subfigure}[b]{0.24\textwidth}
        \centering
\includegraphics[width=1\textwidth]{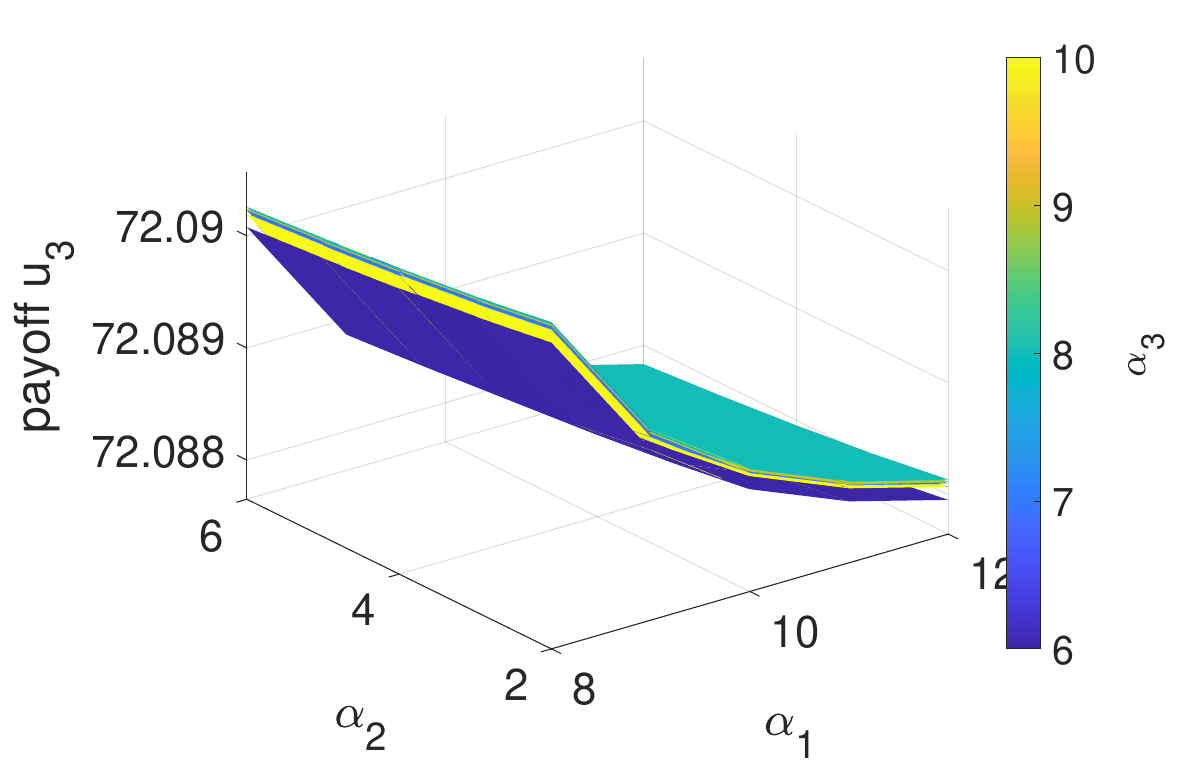}
        \label{Fig.sub.13}
         \caption{}
    \end{subfigure}
    \begin{subfigure}[b]{0.24\textwidth}
        \centering
\includegraphics[width=1\textwidth]{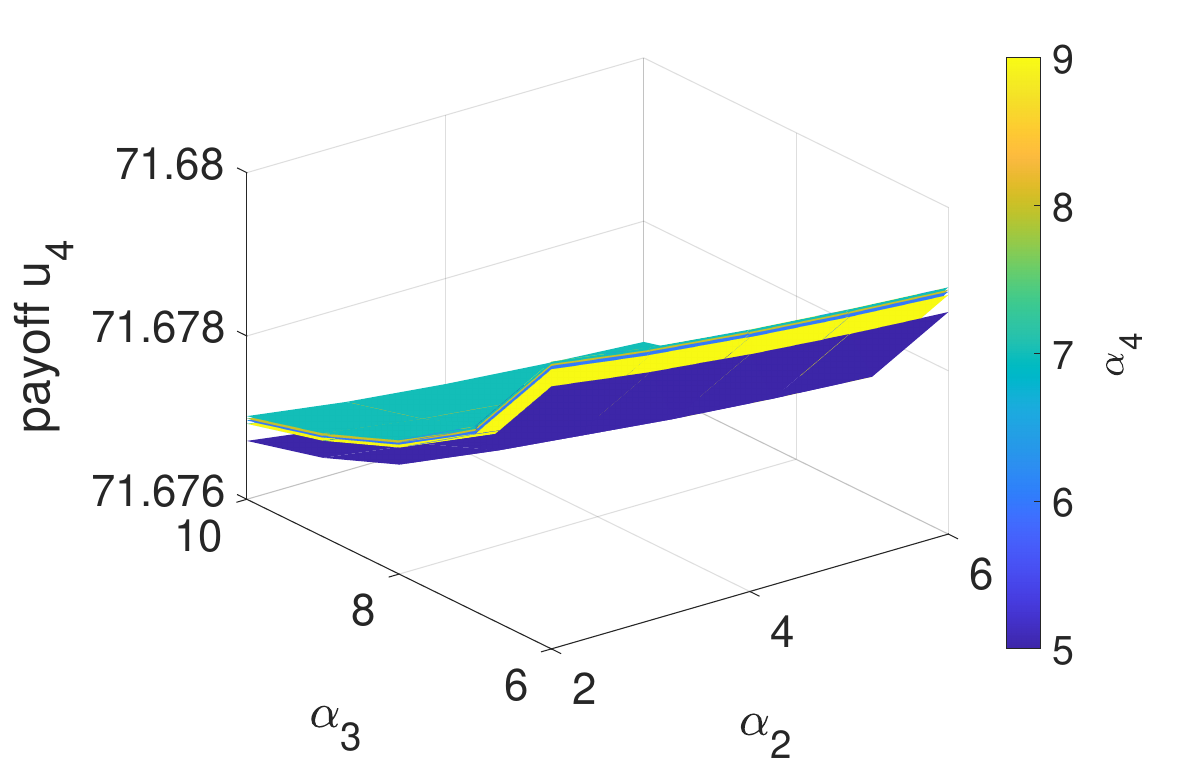}
 \caption{}
        \label{Fig.sub.14}
    \end{subfigure}
\caption{ The payoffs for each agent under various profiles are shown in subfigures (a)–(d), corresponding to the payoffs of agents 1–4, respectively. Each subfigure illustrates five layers, each represented by a different colour, where each colour denotes a specific parameter \(\alpha_i\) of agent \(i\). Within each layer, the plot depicts how agent \(i\)'s payoff varies as a result of changes in the strategies of the other agents. The first three subfigures present the scenario with a fixed \(\alpha_4 = 5\) while the last subfigure presents the payoff of agent 4 with a fixed \(\alpha_1=8\).}
\label{fig7}
\end{figure}

\subsection{TISD manipulation strategies}
In this example, we examine a scenario where all agents employ TISD manipulation strategies. Specifically, we assume that agent \(i\)'s parameter \(\alpha_i\) in sequence \(j\neq i\) is selected based on the following formula:
\begin{align}
    \alpha_i(j)=\alpha_t(i)+\epsilon(i,j),\,i=1,\dots,4, j=1,\dots,4
\end{align}
where \(\alpha_t=[10, 4, 8, 7]\) is the strategies for the social sequence and \(\epsilon(i,j)\) are drawn uniformly from a specified range.

\begin{figure}[H]
\centerline{\includegraphics[width=\columnwidth]{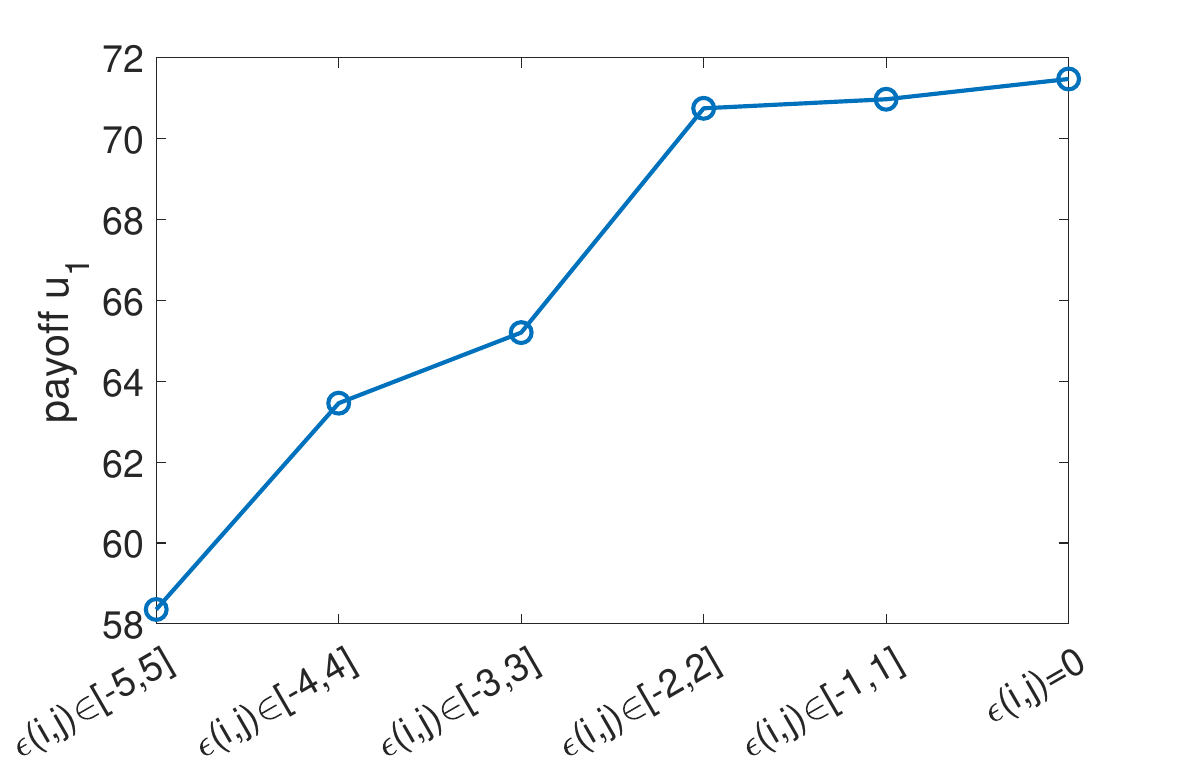}}
\caption{{The payoff of agent 1 vs the range of \(\epsilon(i,j), \forall i,j\in\mathcal{N}\).}}
\label{fig8}
\end{figure}

We adopt a distributed cutting plane algorithm proposed by \cite{zhong2025distributed} with a tolerance \(0.1\) to solve the optimisation problem in each sequence. The total iteration step of the distributed algorithm is \(k_f=300\) and the central unit starts filtering the gradient information from \(k_s=296\). Figure \ref{fig8} shows how agent 1's total cost changes as the range from which \(\epsilon(i,j)\) is drawn varies. The simulation results indicate that adopting the TISD strategy does not provide any advantage to the agent. Notably, when \(\epsilon(i,j)=0\), agent 1 is behaving faithfully. 

\begin{figure}[H]
\centerline{\includegraphics[width=1\columnwidth]{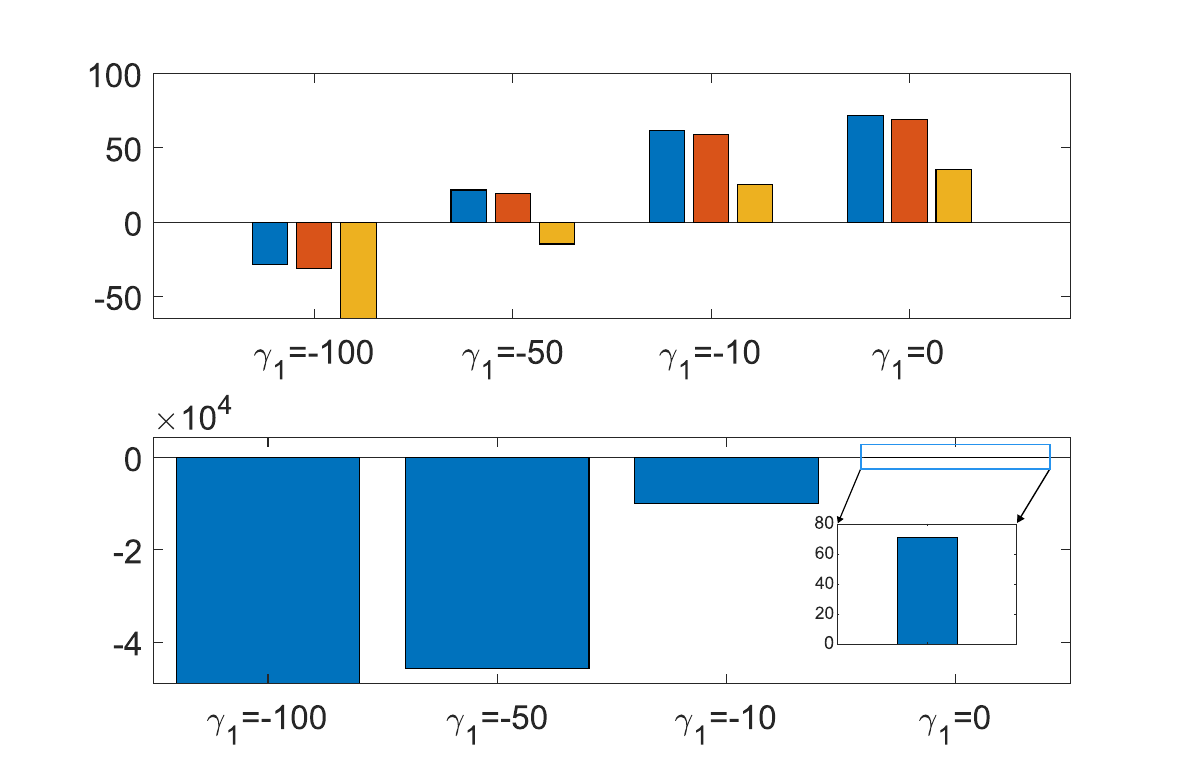}}
\caption{{Agents' payoff are plotted against \(\gamma_1\). The upper plot illustrates the payoffs of agents 2, 3, and 4 (represented by blue, red and yellow). As agent 1 becomes increasingly malicious, their payoffs decrease. The lower plot depicts the decrease in agent 1's payoff as his maliciousness intensifies. Notably, agent 1's payoff decreases at a faster rate than that of the other agents, indicating that being malicious is more harmful to agent 1 than to the others in the DeVCG-G game.}}
\label{fig9}
\end{figure}

We then uniformly select \(\epsilon(i,j)\in[-2,2]\) and let \(\gamma_1\) that represents the maliciousness of agent 1 be a negative value. Figure \ref{fig9} shows how each agent's payoff change with the decrease of \(\gamma_1\).
As shown in the figure, the proposed DeVCG-G mechanism aligns the payoff of the malicious agent (agent 1) with its maliciousness. 
The greater the potential harm an agent can inflict on others, the lower the payoff he will incur.

\section{Future Work and Conclusion}
Apart from adopting time-invariant manipulations, agents may employ time-varying manipulation strategies that allow them to observe their neighbours' actions at each negotiation round. Unlike time-invariant manipulations which are determined before the game begins, time-varying manipulations allow agents to adaptively change their evaluation functions. Initially, each agent selects their evaluation functions individually. From the second step onward, agents observe the actions of their neighbours before determining their subsequent evaluation functions. These evaluation functions are thus informed by historical information across all preceding rounds, leading to coupling among agents' evaluation functions. Future work will focus on modelling this multi-stage game and designing dynamic mechanisms.

To summarise, we have designed a distributed mechanism that can be applied to a set of strategic and malicious agents to solve a social optimisation problem. By utilising the payment rule, which is contributed by the output information from a gradient filter, we proved that the DeVCG-G mechanism is truthful and asymptotically efficient if agents adopt time-invariant manipulation strategies. 


\bibliography{ifacconf}             
 \newpage
\end{document}